\documentclass[aps,prc,twocolumn,showpacs,superscriptaddress,longbibliography,nofootinbib,floatfix,10pt]{revtex4-2}
\usepackage{dcolumn}
\usepackage{amsmath,amssymb,bm,mathrsfs,braket}
\usepackage{multirow,graphicx,longtable,booktabs}
\usepackage[english]{babel}
\usepackage{xcolor}
\usepackage{fix-cm}
\usepackage{mathptmx}
\usepackage[T1]{fontenc}
\usepackage{epstopdf}
\usepackage{soul}

\makeatletter
\def\NAT@def@citea{\def\@citea{\NAT@separator}}
\makeatother
\usepackage{CJKutf8}

\usepackage[colorlinks,citecolor=blue,linkcolor=blue,anchorcolor=blue,
filecolor=blue,urlcolor=blue]{hyperref}

\begin{document}
\begin{CJK*}{UTF8}{}
\hyphenpenalty=8000
\tolerance=1000

\title{Role of the tensor force in induced fission of $^{240}\mathrm{Pu}$}
\author{Yun Huang~\CJKfamily{gbsn}(黄云)}
\affiliation{School of Nuclear Science and Technology, University of Chinese Academy of Sciences, Beijing 100049, China}
\author{Xiang-Xiang Sun~\CJKfamily{gbsn}(孙向向)}
\affiliation{School of Nuclear Science and Technology, University of Chinese Academy of Sciences, Beijing 100049, China}
\affiliation{Institut für Kernphysik, Institute for Advanced Simulation and Jülich Center for Hadron Physics, Forschungszentrum Jülich, D-52425 Jülich, Germany}
\author{Lu Guo~\CJKfamily{gbsn}(郭璐)} \email{luguo@ucas.ac.cn}
\affiliation{School of Nuclear Science and Technology, University of Chinese Academy of Sciences, Beijing 100049, China}
\affiliation{Institute of Theoretical Physics, Chinese Academy of Sciences, Beijing 100190, China}

\date{\today}
		
\begin{abstract}
\edef\oldrightskip{\the\rightskip}
\begin{description}
\rightskip\oldrightskip\relax
\setlength{\parskip}{0pt}
\item[Background] 
Nuclear fission is a large-amplitude collective motion of quantum many-body systems. 
A microscopic description of fission remains a challenge for nuclear theories. 
The tensor force, a key ingredient of the nucleon-nucleon interaction, has attracted great interest in nuclear physics.
Its crucial effects have been revealed in many aspects of both nuclear structure and low-energy heavy-ion reactions. 
However, the role of tensor force in fission is still not well known.

\item[Purpose] 
We study how the tensor force affects the static fission pathway, the dynamical evolution from saddle to scission points, the mass and charge distributions, and the total kinetic energies (TKEs) of fission fragments of the induced fission of $^{240}\mathrm{Pu}$.

\item[Methods] 
Based on the Skyrme density functionals SLy5 and SLy5t, the constrained Hartree-Fock (CHF) with the BCS pairing have been applied to generate a two-dimensional potential energy surface (PES) by constraining quadrupole and octupole mass moments. We then use the time-dependent Hartree-Fock (TDHF) with double particle number projection (PNP) for dynamical calculations.
The TDHF approach is used to simulate the dynamical process until the formation of fission fragments and the properties of fragments can be obtained, in particular their mass, charge, and TKE. 
The double PNP is used to calculate the mass and charge distributions of fragments from each single TDHF fission trajectory,
and then a Gaussian kernel estimation is introduced to give a proper description of fragment distributions.
The changes caused by the tensor force are analyzed by comparing the SLy5 calculations with those of SLy5t.

\item[Results]
We find that the tensor force affects the height of fission barriers and the double-humped structure of fission path of $^{240}\mathrm{Pu}$. 
On the PES, the fission valley becomes larger in the $(Q_{20},~Q_{30})$ plane,
 and the triaxial deformation is suppressed around the outer barrier after considering the tensor force.
Incorporating the tensor force into the dynamical process enhances the difference in shape evolution between two asymmetric channels.
More interestingly, the charge distribution from TDHF with double PNP calculations shows a strong odd-even effect after including the tensor force. 
In addition, the total mass and charge distributions of fission fragments show a slight shift towards larger asymmetry and are more consistent with the experiments when tensor components are included.
We also find that the tensor force enhances energy gaps of the deformed shells for heavy fragments.
Moreover, the TKEs of fragments are in accord with the experiments, 
and after incorporating the tensor force, we observe a higher concentration of the calculated TKEs for heavy fragments at $Z=52$ and $Z=56$.

\item[Conclusions] 
The tensor force plays a role in both static and dynamical processes in nuclear fission, 
improving the accuracy of theoretical descriptions within the current framework.
Our calculations have shown that the fission-barrier height with SLy5t is closer to the empirical value than that of SLy5.
Additionally, the inclusion of the tensor force not only manifests the odd-even effect in charge distribution but also improves the description of total mass and charge distributions.

\end{description}
\end{abstract}
\maketitle
\end{CJK*}
	
\section{Introduction}
Fission is a fundamental nuclear decay and has been of great interest in nuclear physics since its discovery in 1938 \cite{Hahn1939_DN27-11}. 
Many scientific disciplines, including the production of superheavy elements \cite{Hofmann2000_RMP72-733}, the rapid-neutron capture process of nucleosynthesis \cite{Erler2012_PRC85-025802}, and the applications of nuclear technology, depend on the understanding of fission mechanism. 
Nuclear fission is a particularly complex non-equilibrium process due to the transitions between the motion of the mother nucleus and the fast evolution when the neck ruptures and fission fragments are produced \cite{Krappe2012,Schunck2016_RPP79-116301,Simenel2018_PPNP103-19,Bender2020_JPG47-113002,Schunck2022_PPNP125-103963}.
Theoretically, the induced fission process is usually divided into two phases \cite{Simenel2014_PRC89-031601}.
In the first stage, the ground state transits to a configuration that crosses the fission barrier, usually regarded as an adiabatic process.
After the system crosses the fission barrier, it quickly evolves to the scission point and separates into two or three fragments.
Nowadays, almost all the microscopic models of fission are based on the mean-field approximation \cite{Schunck2016_RPP79-116301}.
Determining a multi-dimensional PES is the first step by solving the Hartree-Fock (HF) or the Hartree-Fock-Bogoliubov (HFB) equation.
For the second step, fission dynamics can be achieved through two classes of microscopic methods: the time-dependent generator coordinate method (TDGCM) \cite{Regnier2016_PRC93-054611,Zhao2019_PRC99-054613,Zhao2019_PRC99-014618,Zhao2020_PRC101-064605,Verriere2020_FP8-00233} and the time-dependent density functional theory (TDDFT) \cite{Runge1984_PRL52-997,Simenel2014_PRC89-031601,Scamps2018_Nature564-382,Bulgac2021_PRL126-142502},
which adopt the effective nucleon-nucleon interaction as its only input. 

The tensor force, as an important ingredient of the nucleon-nucleon interaction, provides a strong central attraction in the isospin zero channel, and thus explains well the properties of the deuteron such as its binding energy and non-zero electric quadrupole moment \cite{Rarita1941_PR59-436,Gerjuoy1942_PR61-138}. 
Over the years, the role of tensor force has been actively investigated in nuclear structure.
The tensor force plays a significant role in many exotic nuclear structures, such as the shell evolution of nuclei far from the stability line \cite{Otsuka2005_PRL95-232502,Otsuka2006_PRL97-162501,Sagawa2014_PPNP76-76}, the spin-orbit splitting \cite{Colo2007_PLB646-227}, Gamow-Teller transitions, and charge-exchange spin-dipole excitations \cite{Bai2009_PLB675-28,Bai2010_PRL105-072501,Bai2011_PRC83-054316}.
Due to the tensor force not being included in most studies of nuclear reactions, its influence on nuclear dynamics has recently been a topic of interest.
It has been shown that the tensor force affects the dynamical dissipation in heavy-ion collisions \cite{Dai2014_SCPMA57-1618,Stevenson2016_PRC93-054617}, the potential barrier,
and fusion cross sections \cite{Guo2018_PRC98-064607,Guo2018_PLB782-401,Godbey2019_PRC100-054612,Sun2022_PRC105-034601,Sun2022_CTP74-097302}.
A recent study has also shown that the tensor force strongly affects the evolution of dynamical shell effects in quasifission reactions \cite{Li2022_PLB833-137349}. 
Therefore, the study of the influence of tensor force is required to understand the relation between nuclear structure and microscopic dynamics.
 
Fission is one of the best candidates for revealing the effects of nuclear structure on microscopic dynamics.
However, the role of tensor force in nuclear fission is poorly understand as the tensor force has been neglected in most calculations.
As far as we know, only Ref. \cite{Bernard2020_PRC101-044615} has initially investigated the effect of the tensor force on static fission paths by using HFB calculations with the Gogny forces. 
They found that including the tensor terms can change the barrier height non-negligibly and the topology of the PES, thus accounting for the new compact symmetric fission mode of the neutron-deficient thorium isotopes observed in experiments \cite{Chatillon2019_PRC99-054628}. 
This means that the tensor force is essential for describing the fission of actinide nuclei.
Using the HFB approach, the effects of tensor force on fission are studied only in a static way.
Nuclear fission is definitely a dynamical process so it is meaningful and necessary to study the dynamical effects of the tensor force in fission by using time-dependent approaches.

Up to now, TDDFT has been widely used to describe the induced fission dynamics in a non-adiabatic way.
The roles of shell, pairing correlations, collective excitation modes \cite{Scamps2015_PRC92-011602,Goddard2015_PRC92-054610,Goddard2016_PRC93-014620,Scamps2018_Nature564-382,Qiang2021_PRC103-L031304,Ren2022_PRL128-172501}, the excitation energy sharing mechanism between fission fragments \cite{Bulgac2016_PRL116-122504,Bulgac2019_PRC100-034615,Bulgac2020_PRC102-044609}, and the intrinsic spins of fragments \cite{Bulgac2021_PRL126-142502,Bulgac2021-PRL128-022501} have been investigated in recent years.
However, the absence of the tensor force in these works leads to that the effects of the tensor force on fission are not yet discussed within the framework of TDDFT.
The aim of the present work is to study the effects of tensor force on the induced fission of the benchmark case of $^{240}\mathrm{Pu}$ by using the TDHF approach with double PNP.
In our previous work \cite{Huang2024_EPJA60-100}, we have studied the induced fission process of $^{240}\mathrm{Pu}$ with SLy5 effective interaction without tensor forces, and have shown a reasonable way for a good description of the mass and charge distribution of fission fragments.
As we know, SLy5 and SLy5t are good candidates for investigating the tensor effect. The tensor forces can be clarified by comparing the results with SLy5 and SLy5t, as SLy5t is an extension of SLy5 that includes the tensor terms perturbatively.
Hence, we adopt the SLy5t effective interaction and compare the results in the present work with our previous studies with SLy5 to reveal the influence of tensor forces on both static and dynamical processes in nuclear fission. 
In this work, the effects of tensor force on the static fission pathway and the topology of PES are studied based on the CHF+BCS method.
Then, by implementing the double PNP technique  \cite{Scamps2013_PRC87-014605,Scamps2015_PRC92-011602,Verriere2019_PRC100-024612} in TDHF, the influences from the tensor force on dynamical process and properties of fission fragments are discussed.

This article is organized as follows. In Sec.~\ref{theory}, we describe the theoretical framework. Section~\ref{results} presents the systematic investigates of the impact of the tensor force on the PES and the microscopic mechanism of fission dynamics, respectively. Finally, we summarize our work and give a brief perspective in Sec.~\ref{summary}.

\section{Theoretical framework}
\label{theory}
In this section, we show the adopted effective nucleon-nucleon interaction, i.e., the Skyrme effective interaction and its tensor components. 
Then, brief introductions of the CHF+BCS and the TDHF are given. 
The CHF+BCS is used to generate a multi-dimensional PES by constraining the nucleus shapes and the TDHF can be used to simulate a single fission event in real-time.
Finally, the double PNP technique is presented to get the mass and charge distributions of fission fragments.

\subsection{Full Skyrme effective interaction}
We use the Skyrme effective interaction in our calculation and focus on the effects of tensor force. The tensor terms are written as \cite{Skyrme1958_NP9-615,Stancu1977_PLB68-108}
\begin{equation}
	\begin{split}
		v_{T}=&\frac{t_{\mathrm{e}}}{2}\{[3(\sigma_{1}\cdot~\bm{k}')(\sigma_{2}\cdot~\bm{k}')
		-(\sigma_{1}\cdot\sigma_{2})\bm{k}'^2]\delta(\bm{r}_{1}-\bm{r}_{2})\\
		&+\delta(\bm{r}_{1}-\bm{r}_{2})[3(\sigma_{1}\cdot\bm{k})(\sigma_{2}\cdot~\bm{k})-(\sigma_{1}\cdot\sigma_{2})\bm{k}^{2}]\}\\
		&+t_{\mathrm{o}}[3(\sigma_{1}\cdot~\bm{k}')\delta(\bm{r}_{1}-\bm{r}_{2})(\sigma_{2}\cdot~\bm{k})\\
		&-(\sigma_{1}\cdot~\sigma_{2})\bm{k}'\delta(\bm{r}_{1}-\bm{r}_{2})\bm{k}],
	\end{split}
\end{equation}
where the momentum operator $\bm{k}=\frac{1}{2i}(\nabla_{1}-\nabla_{2})$ acts on the right and $\bm{k}'=-\frac{1}{2i}(\nabla'_{1}-\nabla'_{2})$ for the left. The coupling constants $t_{\mathrm{e}}$ and $t_{\mathrm{o}}$ denote the strengths of triplet-even and triplet-odd tensor interactions, respectively. 

In the Skyrme Hatree-Fock approach, the total energy of the system is expressed in terms of various densities
\begin{equation}
	E=\int d^{3} r \mathscr{H}\left(\rho, \tau, \mathbf{j}, \mathbf{s}, \mathbf{T}, \mathbf{F}, J_{\mu \nu} ; \mathbf{r}\right),
\end{equation}
with the density $\rho$, kinetic energy density $\tau$, current density $\mathbf{j}$, spin density $\mathbf{s}$, spin-kinetic density $\mathbf{T}$, tensor-kinetic density $\mathbf{F}$, and spin-current pseudotensor density $J$ \cite{Engel1975_NPA249-215,Perlinska2004_PRC69-014316,Lesinski2007_PRC76-014312,Sagawa2014_PPNP76-76}. With the inclusion of central, spin-orbit and tensor forces, the full version of Skyrme energy density functional is expressed as
\begin{equation}\label{H}
\begin{aligned}
	\mathscr{H}=& \mathscr{H}_{0}+\sum_{\mathrm{t}=0,1}\left\{A_{\mathrm{t}}^{\mathrm{s}} \mathbf{s}_{\mathrm{t}}^{2}+\left(A_{\mathrm{t}}^{\Delta s}+B_{\mathrm{t}}^{\Delta s}\right) \mathbf{s}_{\mathrm{t}} \cdot \Delta \mathbf{s}_{\mathrm{t}}+B_{\mathrm{t}}^{\nabla s}\left(\nabla \cdot \mathbf{s}_{\mathrm{t}}\right)^{2}\right.\\
	&+B_{\mathrm{t}}^{F}\left(\mathbf{s}_{\mathrm{t}} \cdot \mathbf{F}_{\mathrm{t}}-\frac{1}{2}\left(\sum_{\mu=x}^{z} J_{\mathrm{t}, \mu \mu}\right)^{2}-\frac{1}{2} \sum_{\mu, \nu=x}^{z} J_{\mathrm{t}, \mu \nu} J_{\mathrm{t}, \nu \mu}\right) \\
	&\left.+\left(A_{\mathrm{t}}^{\mathrm{T}}+B_{\mathrm{t}}^{\mathrm{T}}\right)\left(\mathbf{s}_{\mathrm{t}} \cdot \mathbf{T}_{\mathrm{t}}-\sum_{\mu, \nu=x}^{z} J_{\mathrm{t}, \mu \nu} J_{\mathrm{t}, \mu \nu}\right)\right\},
\end{aligned}
\end{equation}
where the $\mathcal{H}_{0}$ is the functional in the TDHF program \texttt{Sky3D} \cite{Maruhn2014_CPC185-2195}
\begin{equation}
	\begin{aligned}
		\mathcal{H}_{0} &=\sum_{\mathrm{t}=0,1}\left\{A_{\mathrm{t}}^{\rho} \rho_{\mathrm{t}}^{2}+A_{\mathrm{t}}^{\Delta \rho} \rho_{\mathrm{t}} \Delta \rho_{\mathrm{t}}+A_{\mathrm{t}}^{\tau}\left(\rho_{\mathrm{t}} \tau_{\mathrm{t}}-\mathbf{j}_{\mathrm{t}}^{2}\right)\right.\\
		&\left.+A_{\mathrm{t}}^{\nabla J} \rho_{\mathrm{t}} \nabla \cdot \mathbf{J}_{\mathrm{t}}+A_{\mathrm{t}}^{\nabla J} \mathbf{s}_{\mathrm{t}} \cdot \nabla \times \mathbf{j}_{\mathrm{t}}\right\}.
	\end{aligned}
\end{equation}
Herein, the coupling constants $A$ and $B$ are defined in Refs. \cite{Lesinski2007_PRC76-014312,Davesne2009_PRC80-024314}. 
Recently, all the terms in Eq. (\ref{H}) have been successfully incorporated into the TDHF calculations by our group \cite{Guo2018_PLB782-401,Guo2018_PRC98-064607,Sun2022_CTP74-097302,Huang2024_EPJA60-100}. Namely, all the time-even and time-odd terms are implemented numerically in the mean-field Hamiltonians of the HF, CHF, and TDHF approaches.
As pointed out in Refs. \cite{Stevenson2016_PRC93-054617,Lesinski2007_PRC76-014312,Shi2017_NPR34-01,Guo2018_PRC98-064609,Guo2018_PLB782-401},the terms containing the gradient of spin density may cause spin
instability in both nuclear structure and reaction dynamics, hence
we turn off the terms of $\mathbf{s}_{\mathrm{t}} \cdot \Delta \mathbf{s}_{\mathrm{t}}$ and $\left(\nabla \cdot \mathbf{s}_{\mathrm{t}}\right)^{2}$ in all calculations.
It should be metioned that the same energy density functional is applied for both the static and dynamical calculations.

\subsection{Constrained Hatree-Fock method}
The Skyrme Hartree-Fock theory with the BCS pairing (HF+BCS) is employed to describe the ground state. 
The BCS approach is implemented to take into account the pairing correlations.
To impose constraints on the nucleus, the augmented Lagrangian method \cite{Staszczak2010_TEPJA46-85} has been used based on the HF$+$BCS in three-dimensional coordinate space and the Routhian reads
\begin{equation}\label{CHF}
	E^{\prime}=E_{\text{HF}}+
	\sum_{i}
	\left[\lambda_{i}
	\left(\langle\hat{\mathcal{Q}}_{i}\rangle-\mathcal{Q}_{i}\right)+
	\frac{1}{2}C_{i}\left(\langle\hat{\mathcal{Q}}_{i}\rangle-\mathcal{Q}_{i}\right)^{2}
	\right],
\end{equation}
where $i$ is used to distinguish  different constrained operators, $\lambda_{i}$ is Lagrangian multiplier, $C_{i}$ is the penalty parameter, and $\mathcal{Q}_{i}$  is the desired value of the $\hat{\mathcal{Q}}_i$.
In our calculations, the constrained operators are either the multipole moments or proportional to them. 
The definitions of multipole moment operators are $\hat{Q}_{\lambda m}(\bm{r})\equiv r^{l}Y_{\lambda m}(\theta,\phi)$,
where $Y_{\lambda m}(\theta,\phi)$ are the spherical harmonics and $\theta$ and $\phi$ are the polar and azimuthal angles.
We employ two types of constrained operator $\hat{\mathcal{Q}}_i$.
One is those that describe the center-of-mass and orientations of a nucleus
\begin{equation}
	\begin{split}
		\hat{\mathcal{Q}}_{x}&=\hat{x},~\hat{\mathcal{Q}}_{y}=\hat{y},~\hat{\mathcal{Q}}_{z}=\hat{z},\\
		\hat{\mathcal{Q}}_{xy}&=\hat{x}\hat{y},~\hat{\mathcal{Q}}_{yz}=\hat{y}\hat{z},~\hat{\mathcal{Q}}_{zx}=\hat{z}\hat{x},\\
	\end{split}
\end{equation}
where these operators are proportional to the $\hat{Q}_{10}$, $\mathrm{Re}  [\hat{Q}_{11}]$, $\mathrm{Im}  [\hat{Q}_{11}]$, $	\mathrm{Im}  [\hat{Q}_{22}]$, $\mathrm{Im}  [\hat{Q}_{21}]$ and $\mathrm{Re}  [\hat{Q}_{21}]$, respectively.
The other is the  quardupole and octupole deformations of a nucleus,
\begin{equation}
	\begin{split}
		\hat{\mathcal{Q}}_{20}&= \hat{Q}_{20}\equiv\sqrt{\frac{5}{16\pi}}(2\hat{z}^{2}-\hat{x}^{2}-\hat{y}^{2}),\\
		\hat{\mathcal{Q}}_{30}&=\hat{Q}_{30}\equiv\sqrt{\frac{7}{16\pi}}\hat{z}(2\hat{z}^{2}-3\hat{x}^{2}-3\hat{y}^{2}).\\
	\end{split}
\end{equation}	
In order to undoubtedly determine the shape of the nucleus, we not only fix the center-of-mass of the nucleus at the center of the numerical box ($\hat{\mathcal{Q}}_{x}=\hat{\mathcal{Q}}_{y}=\hat{\mathcal{Q}}_{z}=0$), but also set the principal axes of the nucleus parallel to the $x$, $y$ and $z$- axis ($\hat{\mathcal{Q}}_{xy}=\hat{\mathcal{Q}}_{yz}=\hat{\mathcal{Q}}_{zx}=0$). The constraints on $\hat{\mathcal{Q}}_{20}$ and $\hat{\mathcal{Q}}_{30}$ for describing elongation and reflection asymmetry of a nucleus are performed to get the static PES.
Other shape degrees of freedom are automatically determined in variation calculations.

\subsection{Time-dependent Hartree-Fock approach}
For a given Hamiltonian of a many-body system, the time-dependent action is written as 
\begin{equation}
S=\int_{t_{1}}^{t_{2}} \mathrm{~d} t\left\langle\Psi(\boldsymbol{r}, t)\left|H-\mathrm{i} \hbar \partial_{t}\right| \Psi(\boldsymbol{r}, t)\right\rangle,
\end{equation}
where $\Psi(\bm{r},t)$ is expressed as the time-dependent many-body wave function. Under the mean-field approximation, $\Psi(\bm{r},t)$ is expressed as the single time-dependent Slater determinant constructed by the single-particle states $\phi_{\alpha}(\bm{r},t)$ and reads
\begin{equation}
	\Psi(\bm{r},t)=\frac{1}{\sqrt{A!}}\mathrm{det}\{\phi_{\alpha}(\bm{r},t)\}.
\end{equation}
By taking the variation of action with respect to $\phi_{\alpha}(\bm{r},t)$, one may obtain the equations of motion for single particle states
\begin{equation}
	i\hbar\frac{\partial}{\partial t}\phi_{\alpha}(\bm{r}, t)=\hat{h}\phi_{\alpha}(\bm{r}, t),
\end{equation}
where $\hat{h}$ is the single-particle Hamiltonian and is also time-dependent. The set of nonlinear TDHF equations has been solved on a three-dimensional Cartesian coordinate system without any symmetry restrictions. 
As a powerful tool to describe the non-equilibrium system, the TDHF theory has many applications in fusion reactions \cite{Guo2012_EWC38-09003,Simenel2013_PRC88-024617,Umar2014_PRC89-034611,Dai2014_PRC90-044609,Umar2016_PRC94-024605,Guo2018_PLB782-401,Guo2018_PRC98-064607,Li2019_SCPMA62-122011,Godbey2019_PRC100-054612,Sun2022_PRC105-034601,Sun2022_PRC105-054610,Sun2023_PRC107-064609,Sun2023_PRC107-L011601}, multi-nucleon transfer reactions \cite{Sekizawa2019_FP7-00020,Wu2019_PRC100-014612,Wu2020_SCPMA63-242021,Wu2022_PLB825-136886}, quasifission \cite{Yu2017_SCPMA60-092011,Guo2018_PRC98-064609,Li2019_SCPMA62-122011,Stevenson2019_PPNP104-142,Godbey2020_FP8-00040,Li2022_PLB833-137349}, deep inelastic collisions \cite{Maruhn2006_PRC74-027601,Guo2007_PRC76-014601,Guo2008_PRC77-041301,Dai2014_SCPMA57-1618,Dai2014_PRC90-044609,Stevenson2016_PRC93-054617,Guo2017_EWC163-00021,Umar2017_PRC96-024625} and fission \cite{Simenel2014_PRC89-031601,Scamps2015_PRC92-011602,Goddard2015_PRC92-054610,Goddard2016_PRC93-014620,Bulgac2016_PRL116-122504,Bulgac2019_PRC100-034615,Scamps2018_Nature564-382,Bulgac2021_PRL126-142502,Bulgac2021-PRL128-022501}. 
Furthermore, to study the dynamics of superfluid systems, one of the extensions of TDHF is to include the pairing correlations. 
In this work, we adopt the BCS approach under the frozen occupation approximation (FOA) \cite{Scamps2013_PRC87-014605}.
Although FOA is a simplified approach to consider dynamical pairing correlations, it may provide, to a certain extent, the reasonable interpretations for the physical scenery. \cite{Simenel2014_PRC89-031601,Goddard2015_PRC92-054610,Goddard2016_PRC93-014620,Simenel2018_PPNP103-19}.
As pointed out in Ref. \cite{Tanimura2017_PRL118-152501}, the introduction of quantum fluctuations could wash out the dynamical pairing correlations. This finding reminds us that the differences of dynamical outcomes caused by various treatments of dynamical pairing require further exploration.

\subsection{Particle number projection technique}
Since the nature of TDHF theory is semi-classical, the trajectories of
FFs are classical and the quantum fluctuations are not included. 
Consequently, the TDHF calculation only provides a mean value of the mass and charge of fission fragments, but cannot give the distributions properly.
The PNP technique based on TDHF \cite{Simenel2010_PRL105-192701,Sekizawa2014_PRC90-064614,Wu2019_PRC100-014612} is employed to extract the particle number fluctuations and its application to fission \cite{Scamps2015_PRC92-011602,Bulgac2019_PRC100-034612,Bulgac2021_PRC104-054601} can give the mass and charge distributions of the fragments.
In this technique, the probability $P^{\tau}_{\mathrm{V}}(N)$ to find $N$ particles with isospin $\tau$ in the subspace $\mathrm{V}$ is given by
\begin{equation}\label{1}
	P^{\tau}_{\mathrm{V}}(N)\equiv \left\langle
	\hat{P}^{\tau}_{\mathrm{V}}(N)\right\rangle
	=\frac{1}{2\pi}\int_{0}^{2\pi}d\eta
	\bra{\Psi}e^{i(\hat{N}^{\tau}_{\mathrm{V}}-N)\eta}\ket{\Psi},
\end{equation}
where $\hat{P}^{\tau}_{\mathrm{V}}(N)$ is the projection operator and $\eta$ is the gauge angle.
$|\Psi\rangle$ is the final state in the TDHF calculation and after rotating it in the gauge space, one obtains $\ket{\Psi_\mathrm{V}}\equiv e^{i\hat{N}^{\tau}_{\mathrm{V}}\eta}\ket{\Psi}$.
 For the case including pairing correlations, the overlap $\left\langle \Psi|\Psi_{\mathrm{V}}\right\rangle$ has a sign ambiguity problem \cite{Onishi1966_NP80-367} due to the square root of a complex matrix. 
In this work, we adopt the Pfaffian method \cite{Robledo2009_PRC79-021302,GonzalezBallestero2011_CPC182-2213} to calculate this overlap.

For superfluid fission system, the influence from the superfluidity of mother nucleus is still ignored by the standard PNP technique [cf. Eq. (\ref{1})], which only considers the particle number fluctuation of fragments in the subspace.
The double PNP technique \cite{Scamps2013_PRC87-014605,Scamps2015_PRC92-011602,Verriere2019_PRC100-024612} is utilized to restore the total number of particles in the whole space. 
Herein, the probability of $N$ protons or neutrons in the subspace V is defined as
\begin{equation}\label{2}
	P^{\tau,~\mathrm{Double}}_{\mathrm{V}}(N)=
	\frac{\bra{\Psi}\hat{P}^{\tau}_{\mathrm{V}}(N)\hat{P}^{\tau}(N_{\mathrm{tot}})\ket{\Psi}}{\bra{\Psi}\hat{P}^{\tau}(N_{\mathrm{tot}})\ket{\Psi}},
\end{equation}
where $N_{\mathrm{tot}}$ is the number of neutrons $(\tau=-1)$ or protons $(\tau=1)$ of the mother nucleus.
The integrals over the gauge angles in Eq. (\ref{1}) and Eq. (\ref{2}) are discretized by using the Fomenko method \cite{Fomenko1970_JPA3-8} with 31 mesh points.
With this equation, the probability of proton number and neutron number in region V can be calculated, and then the mass and charge distributions for a fission event can be obtained. 

	For comparison with experimental data, we fold the mass and charge distributions of all selected fission events. Assuming the mother nucleus splits into two fragments, the charge yield $Y_{\mathrm{V}}(Z)$ and the mass yield $Y_{\mathrm{V}}(A)$ in the subspace $\mathrm{V}$ are given by
\begin{equation}
	\begin{split}
		Y_{\mathrm{V}}(Z) &=\sum_{S}w(S)\times P^{\tau=1,~\mathrm{Double}}_{S,\mathrm{V}}(Z),\\
		Y_{\mathrm{V}}(A) &=\sum_{S}w(S)\times P^{\mathrm{Double}}_{S,\mathrm{V}}(A),\\
	\end{split}
\end{equation}
where $w(S)$ denotes the weight of a given fission trajectory $S$ obtained from the TDHF simulation and
the normalization is $\sum_{S} w(S)=1$.	The yield $Y_{\mathrm{V}}(Z)$ and $Y_{\mathrm{V}}(A)$ are normalized to $2$.

\section{Results and discussions}
\label{results}
In this work, we employ two sets of Skyrme effective interaction, SLy5 \cite{Chabanat1998_NPA635-231} and SLy5t \cite{Colo2007_PLB646-227}. 
These two interactions have been used to investigate the tensor effect in many studies of nuclear structure \cite{Otsuka2006_PRL97-162501,Suckling2010_EPL90-12001} and reactions \cite{Guo2018_PLB782-401,Guo2018_PRC98-064607,Godbey2019_PRC100-054612,Sun2022_PRC105-034601,Li2022_PLB833-137349}.
More details of our code and its applications can be found in
Refs. \cite{Dai2014_PRC90-044609,
	Yu2017_SCPMA60-092011,
	Guo2018_PLB782-401,
	Guo2018_PRC98-064609,
	Guo2018_PRC98-064607,
	Li2019_SCPMA62-122011,
	Wu2019_PRC100-014612,
	Godbey2019_PRC100-054612,
	Wu2020_SCPMA63-242021,
	Wu2022_PLB825-136886,
	Sun2022_PRC105-034601,
	Sun2022_PRC105-054610,
	Sun2023_PRC107-L011601,
	Sun2023_PRC107-064609}.
The static HF equation with the BCS pairing is solved in a three-dimensional grid $36\times32\times36$ $\mathrm{fm}^{3}$ and the grid spacing in each direction is taken to be 1 fm. The constrained calculation is carried out in a larger box with $40\times40\times40$ $\mathrm{fm}^{3}$ when generating the PES in the $(Q_{20},~Q_{30})$ plane.
For the pairing correlations, we adopt the volume delta pairing force with defalut pairing strengths $V^{nn}_{0}=288.523$ $\mathrm{MeV}$ $\mathrm{fm}^{3}$ and $V^{pp}_{0}=298.760$ $\mathrm{MeV}$ $\mathrm{fm}^{3}$ for neutrons and protons, respectively \cite{Maruhn2014_CPC185-2195}.
The time-dependent calculations are performed in a box with the size of $48\times48\times160~\mathrm{fm}^{3}$. Fission occurs along the $z$ axis.
The time evolution operator is evaluated by the Taylor series expansion at the eighth order and the time step is taken to be $0.25$ $\mathrm{fm/c}$.
The center-of-mass correction is not taken into account to maintain consistency between static and dynamical calculations.
The technical implementation of current static and dynamical calculations is the same as our previous study \cite{Huang2024_EPJA60-100}.
All these parameters are chosen to guarantee good numerical accuracy for the present studies.

\subsection{Two-dimensional potential energy surface of $^{240}\mathrm{Pu}$}
\label{PES}

\begin{figure}[htbp]
	\includegraphics[width=0.48\textwidth]{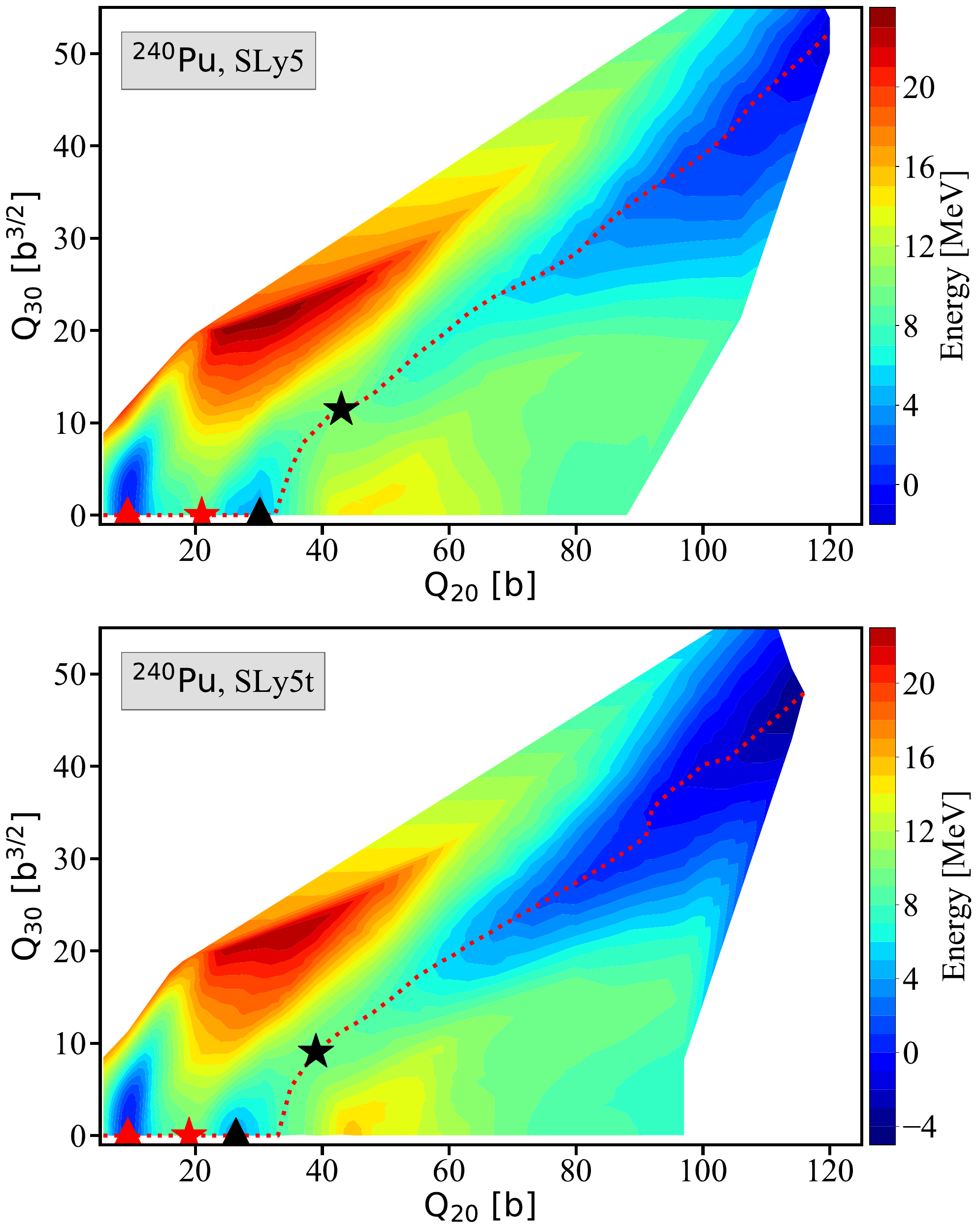}
	\caption{Potential energy surface of $^{240}\mathrm{Pu}$
		in the $(Q_{20},~Q_{30})$ plane calculated with SLy5 (the upper panel) and SLy5t (the bottom panel). The red dotted line shows the static fission pathway. The red triangle, red star, black triangle, and black star denote the ground state, inner barrier, isomeric state, and outer barrier, respectively. Each panel shows the normalized energies relative to their ground state values.}
	\label{Fig:PES}
\end{figure}

\begin{figure}[htbp]
	\includegraphics[width=0.48\textwidth]{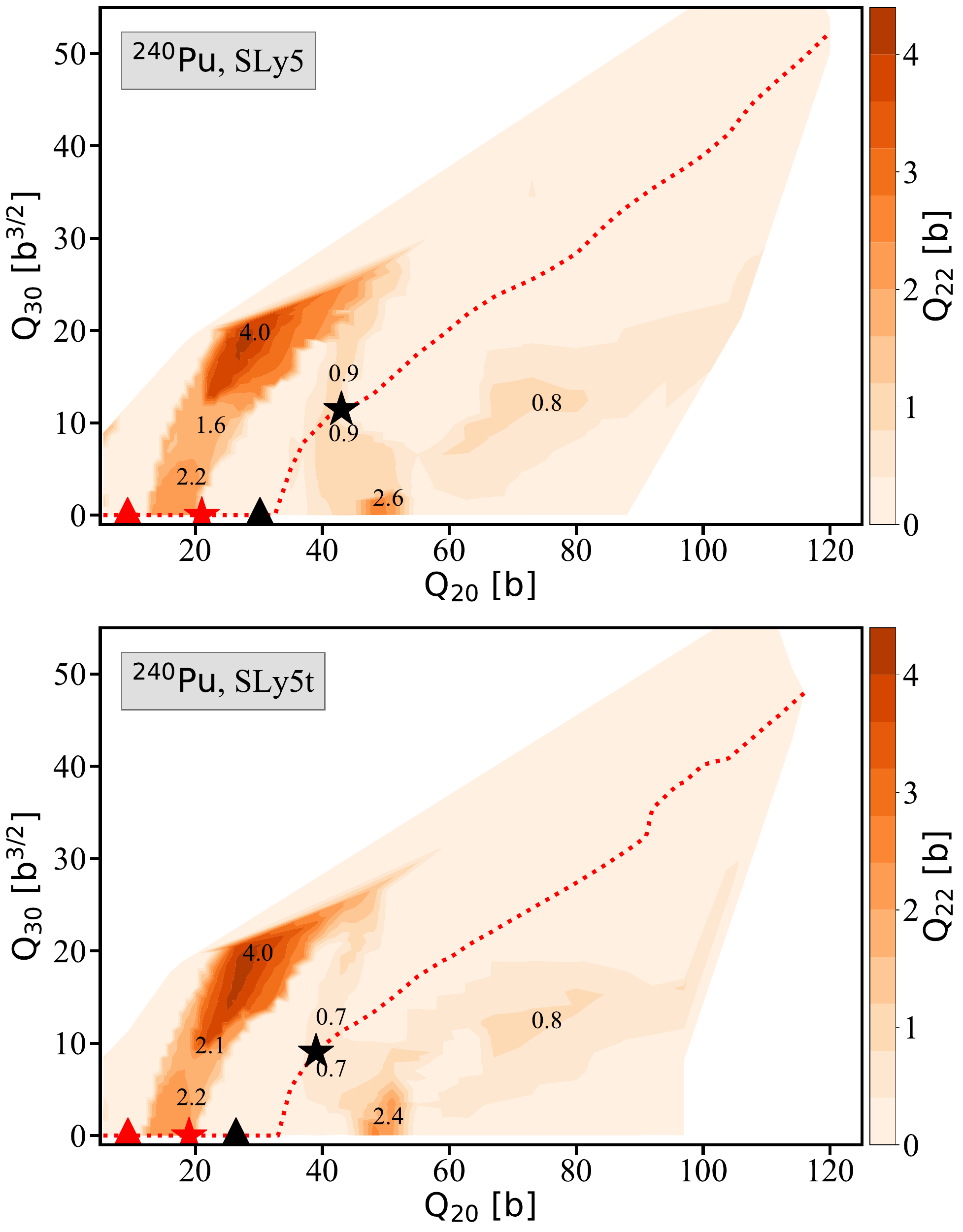}
	\caption{The triaxial deformation of $^{240}\mathrm{Pu}$
		in the $(Q_{20},~Q_{30})$ plane calculated with SLy5 (the upper panel) and SLy5t (the bottom panel).
		The numbers show the values of $Q_{22}$ at selected points.}
	\label{Fig:Q22}
\end{figure}

The calculation of a multi-dimensional PES is the first step in the microscopic study of nuclear fission.
We present a two-dimensional (2D) PES generated by the CHF+BCS method with the constraints of both the quadruple mass moment $Q_{20}$ and octuple mass moment $Q_{30}$.
In Fig. \ref{Fig:PES}, we show the PESs of $^{240}\mathrm{Pu}$ in $(Q_{20},~Q_{30})$ plane using SLy5 and SLy5t effective interactions. 
The red dotted lines present the static fission path, which refers to a minimum energy path on the 2D PES.
The ground state, inner barrier, isomeric state, and outer barrier are denoted by red triangle, red star, black triangle, and black star respectively, which are crucial points to reflect the property of the fission nucleus. 
The double-humped structure of the static fission path can be seen for these two PESs and the mass asymmetric deformation allows for decreasing the outer barrier.
In general, the inclusion of tensor force does not affect the main features of the topology of PES, apart from making the PES flatter slightly.
In the regions near the isomeric state and outer barrier, the PES becomes softer with the tensor force.
In addition, beyond the outer barrier, one can see that the fission valley is apparently enlarged by the tensor force.
Such a broad fission valley may be favored for dynamical evolution.
This is due to the fact that the structure of single-particle levels and their occupation probabilities are different for the cases of SLy5 and SLy5t.

Besides $Q_{20}$ and $Q_{30}$, the triaxial mass moment $Q_{22}$ is also crucial for determining the fission barrier and the PES topology. 
The importance of $Q_{22}$ on the fission barrier has been confirmed by earlier studies \cite{Larsson1972_PLB38-269,Girod1983_PRC27-2317,Rutz1995_NPA590-680,Schunck2014_PRC90-054305}.
We present the triaxial deformation on the 2D PESs in Fig. \ref{Fig:Q22}.
The nonzero triaxial deformation can be found in two regions.
The first one starts from the inner barrier and extends roughly along the direction of the $Q_{30}$ axis up to a very asymmetric shape with $Q_{30} = 20~ \mathrm{b}^{3/2}$.
In both SLy5 and SLy5t cases, the value of $Q_{22}$ along the $Q_{30}$ axis initially decreases and then increases.
The other one is around the outer barrier, and the $Q_{22}$ values are about $0.9~\mathrm{b}$ for the SLy5, and $0.7~\mathrm{b}$ for the SLy5t. 
Comparing the results of two effective interactions, the tensor force has no effect on the triaxial deformation of the former region but it slightly lowers the values of  $Q_{22}$ around the outer barrier. 

\begin{figure}[htbp]
	\includegraphics[width=0.4\textwidth]{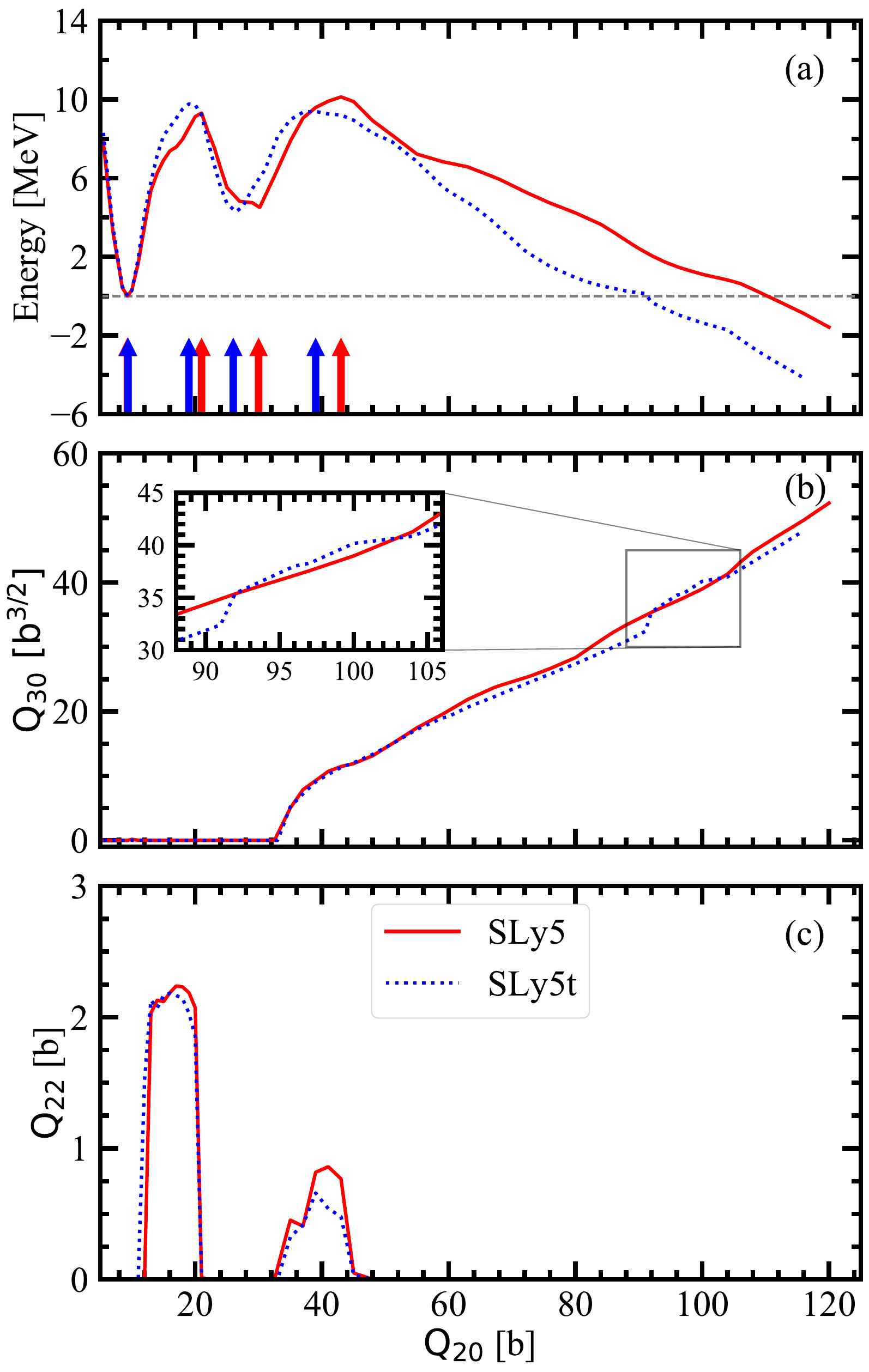}
	\caption{The relative energies (a), octupole deformation (b), and triaxial deformation (c) along the asymmetric fission path as a function of quadrupole mass moment. The red solid curve and the blue dotted curve are the results for the SLy5 and the SLy5t, respectively. In panel (a), the horizontal dashed line marks the position where the relative energy is zero and the up arrows label the locations of the ground state, inner fission barrier, isomeric state, and outer fission barrier, respectively.
		The inset on panel (b) is an enlarged view of the fission pathway in the region $88~\mathrm{b}\leq Q_{20}\leq 106~\mathrm{b}$. }
	\label{Fig:PEC-Q30-Q20}
\end{figure}

To examine the effects of tensor force on PES further, we present the relative energies along the asymmetric fission pathway as functions of $Q_{20}$ in Fig. \ref{Fig:PEC-Q30-Q20} (a), the corresponding evolution of $Q_{30}$ in Fig. \ref{Fig:PEC-Q30-Q20} (b) and $Q_{22}$ in Fig. \ref{Fig:PEC-Q30-Q20} (c). 
In Fig. \ref{Fig:PEC-Q30-Q20} (a), the positions of the ground state, inner barrier, isomeric state, and outer barrier in the fission pathway are indicated by red up arrows for SLy5 and blue up arrows for SLy5t.
It is found that the deformation of ground state does not change and the values of $Q_{20}$ corresponding to the inner barrier, isomeric state, and outer barrier become smaller after considering the tensor force.
The energies and deformation parameters of these configurations are listed in Table \ref{table:property}. 
As for the height of the fission barrier obtained from the 2D PES, including the tensor force results in a higher inner barrier, but lowers the outer barrier. 
In Fig. \ref{Fig:PEC-Q30-Q20} (a), the energy beyond the fission barrier decreases more rapidly in the SLy5t case. The energy difference between SLy5 and SLy5t is about 2.5 MeV when $Q_{20}>67~\mathrm{b}$.

\begin{table*}[ht]
	\centering
	\caption{The energies of the ground state, inner barrier, isomeric state and outer barrier for $^{240}\mathrm{Pu}$ with SLy5 and SLy5t. The quadrupole, triaxial and octupole deformations of these states are also given.}
	\begin{ruledtabular}
		\begin{tabular}{*{9}{c}}
			& \multicolumn{4}{c}{SLy5} &\multicolumn{4}{c}{SLy5t} \\
			& Energy~(MeV) & $Q_{20}~(\mathrm{b})$ & $Q_{22}~(\mathrm{b})$  & $Q_{30}~(\mathrm{b}^{3/2})$ & Energy~(MeV) & $Q_{20}~(\mathrm{b})$ & $Q_{22}~(\mathrm{b})$  & $Q_{30}~(\mathrm{b}^{3/2})$ \\
			\midrule 
			Ground state 	  	& $-1784.83$ & 9.3  & 0.0 & 0.0  & $-1792.24$ & 9.4  & 0.0 & 0.0   \\
			Inner barrier  		& $-1775.52$ & 21.0 & 0.0 & 0.0  & $-1782.47$ & 19.0 & 2.0 & 0.0   \\
			Isomeric state  	& $-1780.32$ & 30.0 & 0.0 & 0.0  & $-1787.95$ & 26.0 & 0.0 & 0.0    \\
			Outer barrier  		& $-1774.70$ & 43.0 & 0.8 & 11.4 & $-1782.85$ & 39.0 & 0.7 & 9.1  \\
		\end{tabular}%
	\end{ruledtabular}
	\label{table:property}
\end{table*}%

\begin{figure*}[htbp]
	\includegraphics[width=1\textwidth]{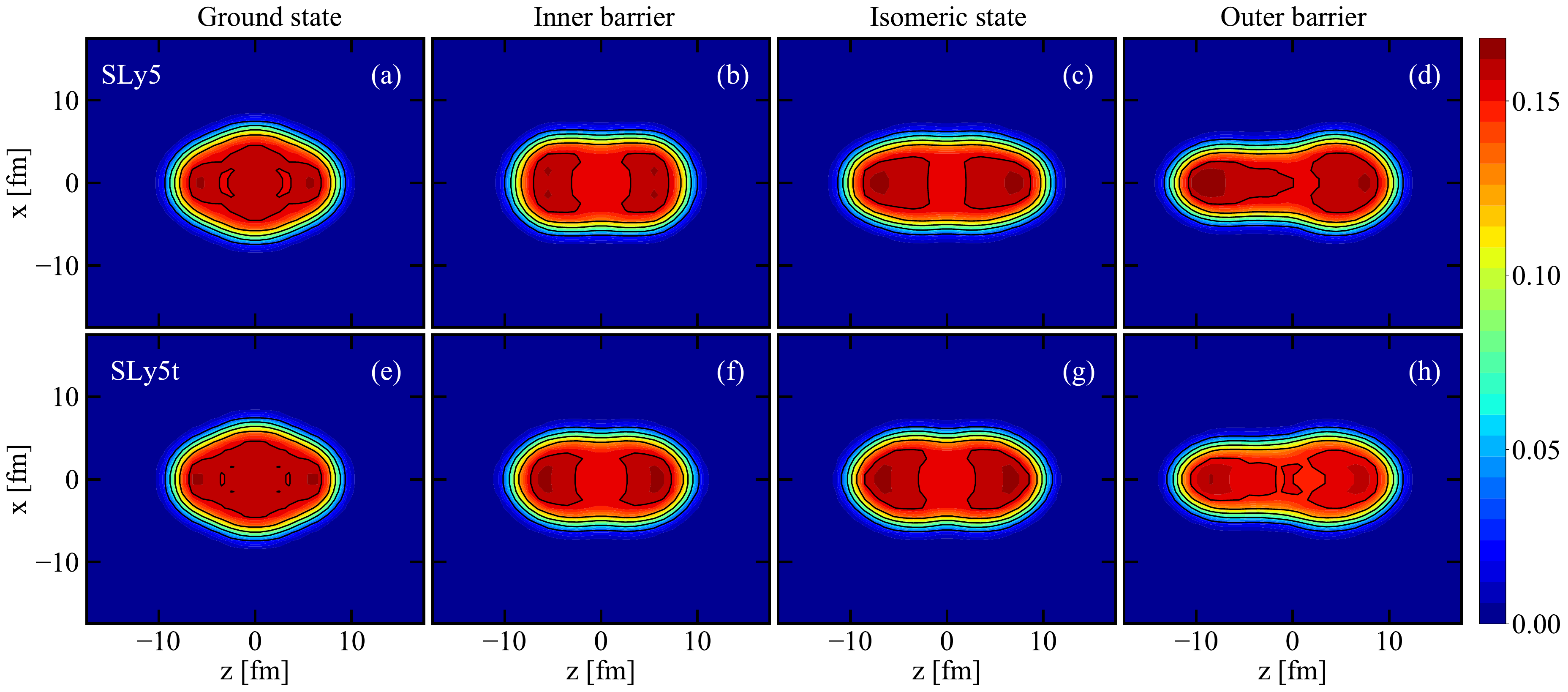}
	\caption{The 2D density distributions for the ground state, inner barrier, isomeric state and outer barrier. The upper panel shows the results with SLy5 and the bottom panel for SLy5t.}
	\label{Fig:multdensity}
\end{figure*}

Fig. \ref{Fig:PEC-Q30-Q20} (b) displays $Q_{30}$ along the asymmetric fission pathway as a function of $Q_{20}$. 
It is noticed that along the fission pathway of SLy5, the $Q_{30}$ plays a role beyond the isomeric state.
Generally speaking, the trend of $Q_{30}$ with respect to $Q_{20}$ in the SLy5t calculations is similar to that of SLy5.
Along the fission path, the values of $Q_{30}$ in SLy5 calculations are always larger than those of SLy5t except the region $88~\mathrm{b} \leq Q_{20} \leq106~\mathrm{b}$, as shown more clearly in the inset of Fig. \ref{Fig:PEC-Q30-Q20} (b).
This behavior is not so obvious in the potential energy and it may be a sign of the complex topology of PES.

Fig. \ref{Fig:PEC-Q30-Q20} (c) shows the $Q_{22}$ on the asymmetric fission pathway of SLy5 and SLy5t. 
We find that the triaxial deformation mainly occurs in two regions, around the inner barrier and the outer barrier. 
The triaxial deformation is important for both the inner and outer barrier, which is consistent with other theoretical models \cite{Lu2012_PRC85-011301,Schunck2014_PRC90-054305}. There is no qualitative difference in the triaxial deformation caused by tensor force along the static fission path.
The $Q_{22}$ for SLy5t calculations around the $Q_{20}=40~\mathrm{b}$ is smaller than that of SLy5.
It is well known that the inner fission barrier is lowered due to the triaxial deformation \cite{Larsson1972_PLB38-269,Girod1983_PRC27-2317,Rutz1995_NPA590-680}. 
The effect of the triaxiality on the outer fission barrier is still an open question and opposite conclusions are obtained by the macroscopic-microscopic model \cite{Jachimowicz2012_PRC85-034305} and the microscopic models \cite{Abusara2012_PRC85-024314,Lu2014_PRC89-014323}. 
In our 2D PES, we found that that triaxial deformation is slightly quenched due to the tensor force thus lowers the outer barrier. A more detailed study of the role of $Q_{22}$ on the outer barrier with or without the tensor force need to construct a three-dimensional PES, but which is beyond the scope of present work.

The values of the energies and deformations of ground state, inner barrier, isomeric state, and outer barrier for SLy5 and SLy5t are listed in Table \ref{table:property}. 
The energies of these states become lower after considering the tensor force.
After considering the center-of-mass correction, the binding energy of ground state changes from 1784.83 to 1803.52 MeV for SLy5, and from 1792.24 to 1811.02 MeV for SLy5t.
Notably, the binding energy calculated with the SLy5t is closer to the empirical value (1813.45 MeV) in AME2020 \cite{Kondev2021_CPC45-030001,Huang2021_CPC45-030002,Wang2021_CPC45-030003}. 
Besides, after including the tensor force, the deformation of the ground state remains unchanged and it locates at around $Q_{20}=9.4~\mathrm{b}$.
Other states have smaller deformations in SLy5t, except for the triaxiality of inner barrier. 
The triaxial deformation of the inner barrier occurs only in the SLy5t case, i.e., the tensor force affects the triaxiality of the inner barrier.
Fig. \ref{Fig:multdensity} shows the 2D density distributions at ground state, inner barrier, isomeric state, and outer barrier. 
It is found that the differences between SLy5 and SLy5t calculations become more evident with the increase of deformation. 

Table \ref{table:fissionbarrier} lists the height of the inner barrier ($\text{B}^{\text{i}}_{\text{f}}$) and the outer barrier ($\text{B}^{\text{o}}_{\text{f}}$), as well as the excitation energy of isomer ($E_{\text{II}}$) in our calculations.  
Our calculations of these three values with or without tensor force overestimate the empirical values taken from Ref. \cite{Moeller2009_PRC79-064304}. 
Similar conclusions are also obtained for actinides from the calculations with Skyrme density functionals given in Refs. \cite{Flocard1974_NPA231-176,Brack1979,Buervenich2004_PRC69-014307}. 
Especially, although the information of barrier of $^{240}\mathrm{Pu}$ has already been considered when determining the SkM* force, the calculations with SkM* still overestimate the empirical values \cite{Goddard2015_PRC92-054610}.
In fact, there are many factors affecting the fission barrier, such as the surface energy coefficient \cite{Jodon2016_PRC94-024335,Ryssens2019_PRC99-044315}, the pairing correlations \cite{Schunck2014_PRC90-054305} and some beyond-mean-field effects \cite{Baran2014_PS89-054002}.
Herein, we focus on the impact of tensor force on the height of fission barrier. 
In SLy5, the inner barrier is lower than the outer barrier.
With the inclusion of the tensor force, the inner-barrier height increases by 0.46 MeV while the outer-barrier height decreases by 0.74 MeV. 
Consequently, in SLy5t, the inner barrier becomes higher than the outer one.
To quantify the change caused by the tensor force, we extract the ratio of the inner-barrier height to the outer-barrier height ($\text{B}^{\text{i}}_{\text{f}}/\text{B}^{\text{o}}_{\text{f}}$), which is listed in the last column of Table \ref{table:fissionbarrier}.
It is noted that the ratio of the SLy5t coincides with the ratio of the empirical values.
To summarize, in the static calculations, one can conclude that the tensor force enlarges the fission valley in the ($Q_{20},~Q_{30}$) plane, lowers the triaxial deformation in the outer barrier region, and strongly influences the height of fission barriers.

\begin{table}[t]
	\centering
	\caption{The inner barrier height ($\text{B}^{\text{i}}_{\text{f}}$), outer barrier height ($\text{B}^{\text{o}}_{\text{f}}$), excitation energy of the isomer ($E_{\text{II}}$) and the ratio of $\text{B}^{\text{i}}_{\text{f}}/\text{B}^{\text{o}}_{\text{f}}$ in our calculations with SLy5 and SLy5t, and the comparison with other models. 
	The results with SkM* taken from Ref. \cite{Goddard2015_PRC92-054610} and the empirical values (``Emp'') from Ref. \cite{Moeller2009_PRC79-064304} are shown for comparison.}
	\resizebox{\linewidth}{!}{       
		\begin{ruledtabular}
			\begin{tabular}{*{5}{c}}
				Force & $\text{B}^{\text{i}}_{\text{f}}$ (MeV) & $\text{B}^{\text{o}}_{\text{f}}$ (MeV) & $E_{\text{II}}$ (MeV) & $\text{B}^{\text{i}}_{\text{f}} / \text{B}^{\text{o}}_{\text{f}}$ \\
				\midrule 
				SLy5  & 9.31 & 10.13 & 4.51 & 0.919 \\
				SLy5t & 9.77 & 9.39 & 4.29 &  1.040  \\
				SkM* \cite{Goddard2015_PRC92-054610} & 8.25 & 7.68 & 3.04 & 1.074 \\
				Emp \cite{Moeller2009_PRC79-064304}   & 6.1$\pm$0.3  & 6.0$\pm$0.5  & 2.1$\pm$0.6  & 1.017$\pm$0.037 \\
			\end{tabular}%
	\end{ruledtabular}}
\label{table:fissionbarrier}
\end{table}%

\subsection{Dynamic evolution}
\label{dynamical}

\begin{figure}[htbp]
	\includegraphics[width=0.48\textwidth]{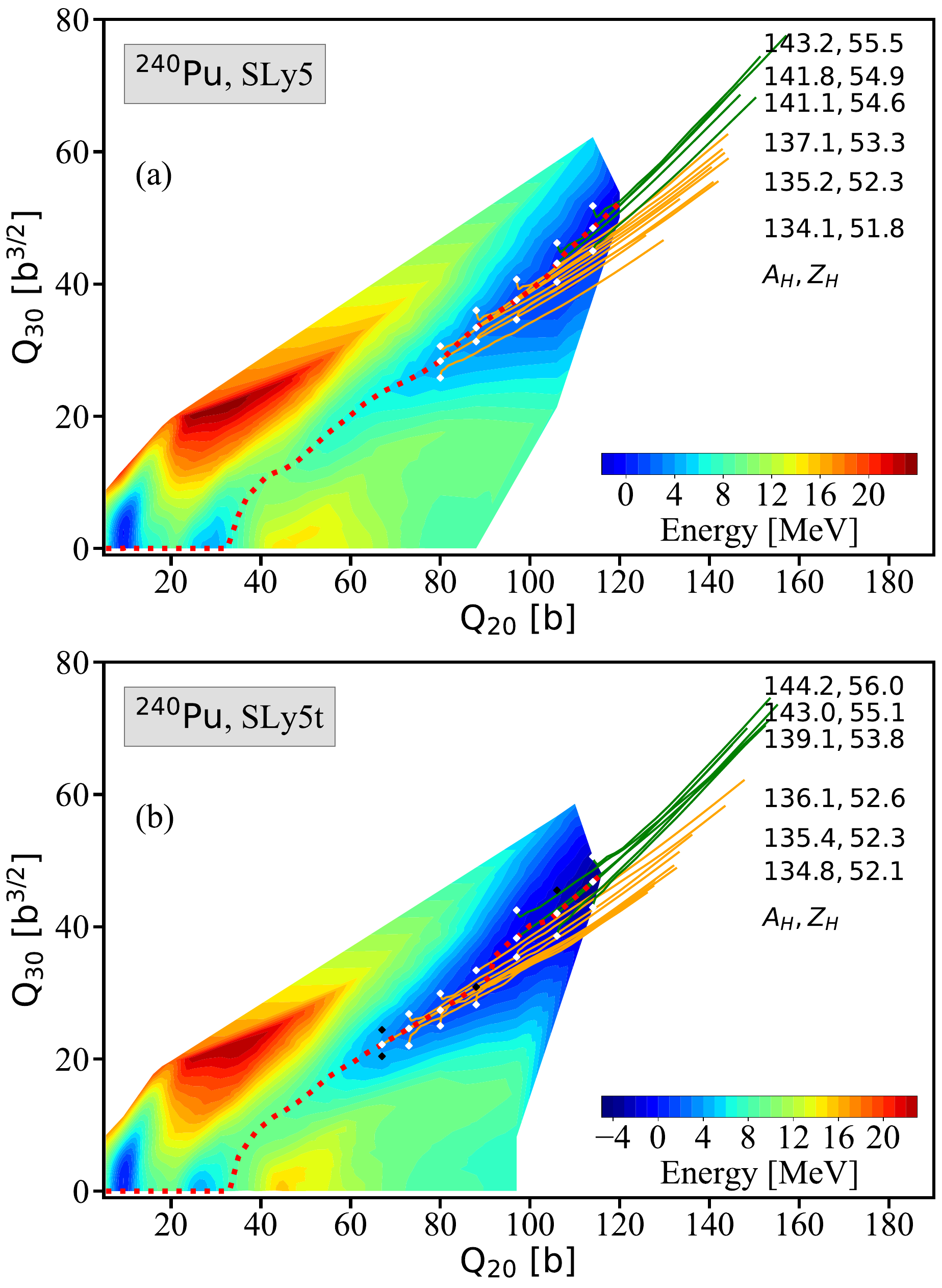}
	\caption{The TDHF fission trajectories of $^{240}\mathrm{Pu}$ with the SLy5 (the upper panel) and SLy5t (the bottom panel) in the space of $(Q_{20},~Q_{30})$. 
		Selected static configurations of TDHF calculations are marked by the diamonds. 
		The initial points that can fission are represented by the white diamonds, while the initial points that cannot fission are represented by the black diamonds.
		The yellow and green lines correspond to two kinds of dynamical pathways. 
		A series of mass $A_{H}$ and charge $Z_{H}$ numbers of heavy fragments are also shown on the right.}
	\label{Fig:TimeEvolution}
\end{figure}

Based on the 2D PES in the $(Q_{20},~Q_{30})$ plane, we choose a series of configurations as the initial points of dynamical simulation.
As pointed out in Ref. \cite{Scamps2015_PRC92-011602}, the TDHF+FOA cannot be used to simulate the evolution from the equilibrium deformation to the saddle point.
Consequently, the initial configuration is usually taken beyond the outer barrier. 
That is to say, there exists a dynamical fission threshold on the static fission pathway \cite{Goddard2015_PRC92-054610}. 
For such a threshold, its inhibition of fission can be explained by the lack of dynamical pairing \cite{Bulgac2016_PRL116-122504,Tanimura2017_PRL118-152501}.
In our calculations, the threshold also exists.
The threshold on the static fission pathway of the SLy5 case is about $Q_{20}=80~\mathrm{b}$, while that of the SLy5t case is $Q_{20}=67~\mathrm{b}$. 
The difference in threshold is likely due to the combination of two factors: different heights of fission barriers and slopes of the static fission pathway beyond the outer barrier. 
In addition, this threshold anomaly also occurs in the $Q_{30}$ direction.
As a result, more non-fission events will be faced when considering a wide range of $Q_{30}$ values.
In this work, a nearly equidistant pattern in both $Q_{20}$ and $Q_{30}$ directions is taken into account.
Specifically, we select seven points with different $Q_{20}$ varying from $67~\mathrm{b}$ to $114~\mathrm{b}$ on the static fission pathway, and then, for each value of $Q_{20}$, additional two points with different $Q_{30}$ are chosen in the fission valley.
This means that a total of 21 states are selected in 2D PES as the initial configurations for dynamical calculations.
According to their respective fission thresholds, 15 states are chosen in the SLy5 case, while 21 states for the case with SLy5t.

\begin{figure*}[htbp]
	\includegraphics[width=0.96\textwidth]{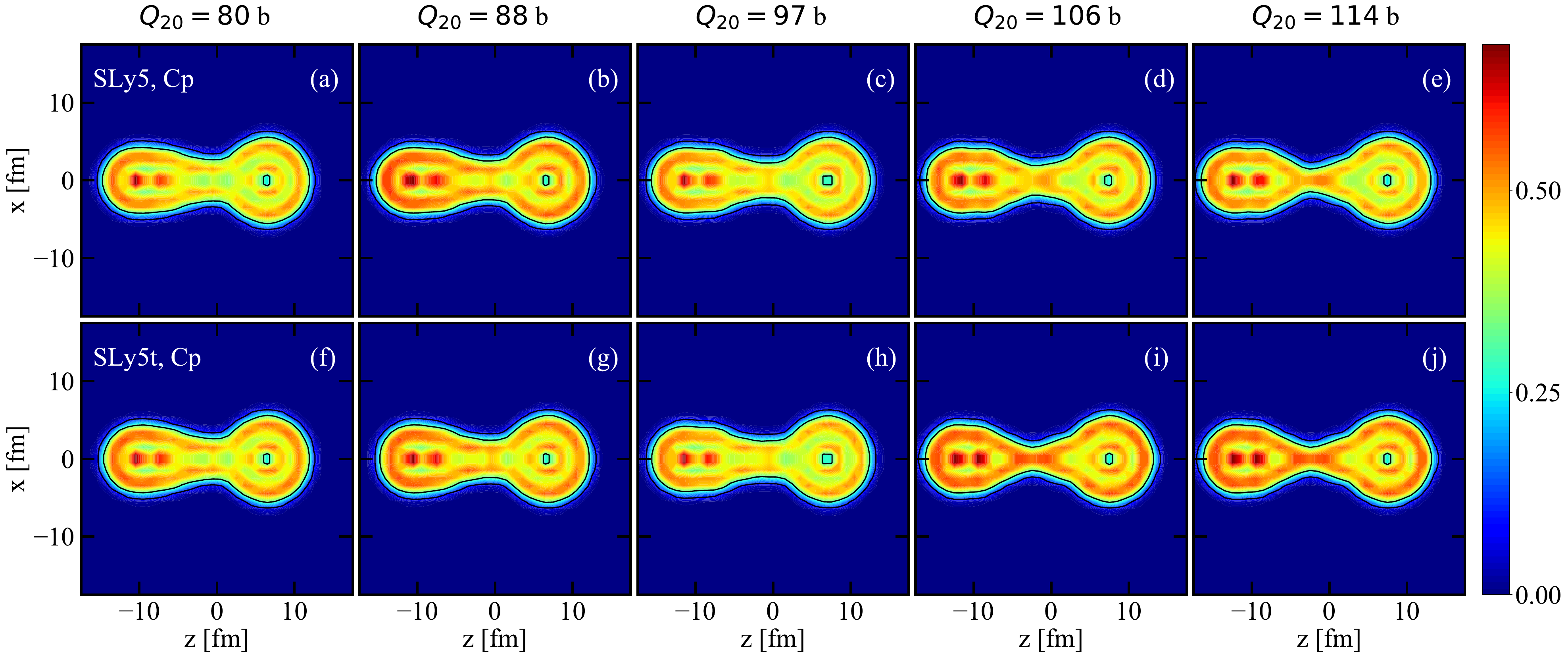}
	\caption{The proton localization functions $C_{p}$ for five configurations along the static fission pathway. The corresponding quadrupole moments are displayed on the top of each column. The upper panel shows the results with SLy5 and the bottom panel for SLy5t.}
	\label{Fig:Cp}
\end{figure*}

With these initial states, we perform the TDHF+FOA calculations and the corresponding dynamical fission trajectories are shown by the solid lines in Fig. \ref{Fig:TimeEvolution} (a) for SLy5 and Fig. \ref{Fig:TimeEvolution} (b) for SLy5t. 
All these initial states are represented by diamonds.
In our calculations, fission happens for some of initial configurations indicated by white diamonds in Fig. \ref{Fig:TimeEvolution} while the black ones are for the cases where fission does not happen.
In the case of SLy5, all 15 initial states can fission and these states have a mean excitation energy of 2.1 MeV with a variance of 3 MeV for $^{240}\mathrm{Pu}$. 
In SLy5t, there are 17 fissioning states corresponding to a mean excitation energy of 0.69 MeV with a variance of 4.6 MeV, and four non-fissioning states.
The most interesting non-fissioning state is $Q_{20}=88~\mathrm{b}$ on the asymmetric fission path, because it locates between two fissioning states.
This might be due to the different behavior of the fission pathway caused by the tensor force, as shown in Fig. \ref{Fig:PEC-Q30-Q20}. 
A deep understanding requires carefully investigating the structure of PES more than two dimensions and corresponding dynamical simulations, which needs enormous computational resources beyond our ability. 

\begin{table}[t]
	\centering
	\caption{The mean value and the standard deviation of the mass number and charge number of heavy fragments for SI and SII channels in both SLy5 and SLy5t cases. }
	\resizebox{\linewidth}{!}{       
		\begin{ruledtabular}
			\begin{tabular}{*{9}{c}}
				\multirow{2}{*}{Force}& \multicolumn{4}{c}{SI channel} &\multicolumn{4}{c}{SII channel} \\
				 & $\bar{A}_{H}$  & $\Delta A_H$ & $\bar{Z}_{H}$  & $\Delta Z_H$ & $\bar{A}_{H}$  & $\Delta A_H$ &
				 $\bar{Z}_{H}$  & $\Delta Z_H$ \\
				\midrule 
				SLy5  & 135.2 & 0.90 & 52.3 & 0.45 & 142.4 & 0.94 & 55.1 & 0.36 \\
				SLy5t & 135.5 & 0.43 & 52.4 & 0.19 & 142.3 &  2.40 & 55.1 & 0.93 \\
			\end{tabular}%
	\end{ruledtabular}}
	\label{table:meanvalue}
\end{table}%

\begin{figure*}[htbp]
	\includegraphics[width=0.96\textwidth]{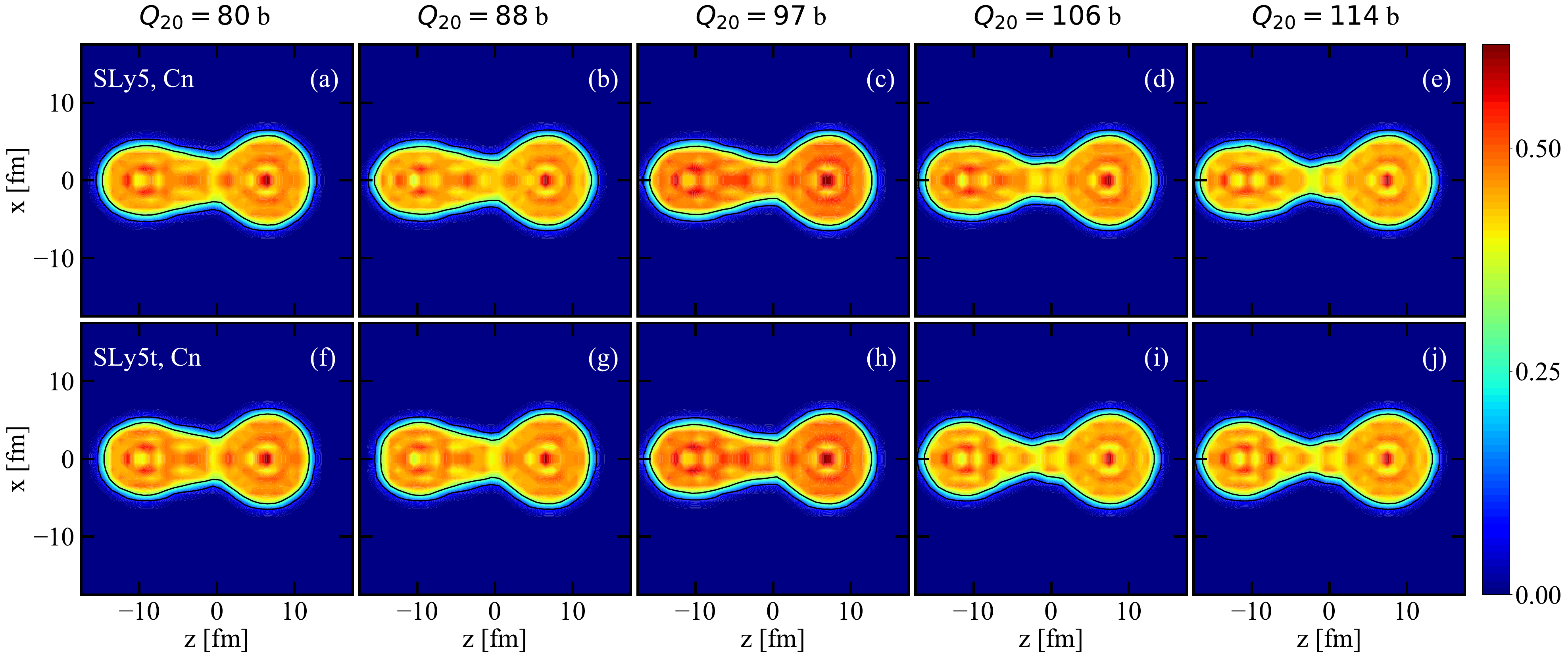}
	\caption{Similar to Fig. \ref{Fig:Cp} but for the neutron localization functions $C_{n}$.}
	\label{Fig:Cn}
\end{figure*}

In Fig \ref{Fig:TimeEvolution}, we can see that in both cases, the octuple moments of all fission trajectories increase with the elongation of the fissioning nucleus.
When the distance between the centers of mass of two primary fission fragments is larger than 30 fm, the dynamical calculation ends and the mass of primary fragments can be directly obtained by integrating one-body local densities in the subspace. 
We also list a series of the mass and charge numbers of heavy fragments in Fig. \ref{Fig:TimeEvolution}, labeled by $A_{H}$ and $Z_{H}$.
In SLy5 and SLy5t cases, the mass and charge numbers of heavy fragments are almost $A_H=134$~--~$144$ and $Z_H=52$~--~$56$.
In terms of $A_H$, there are two kinds of dynamical pathways. 
One is those fission trajectories with $A_H$ close to $135$ marked by yellow lines and the other with $A_H\approx142$ presented by green lines. 
These two types correspond to the SI and SII channels of the Brosa model \cite{Brosa1990_PR197-167}, respectively.  The mean value and the standard deviation of $A_H$ and $Z_H$ for SI and SII channels are listed in Table \ref{table:meanvalue}, which are represented by $\bar{A}_{H}$, $\Delta A_H$, $\bar{Z}_{H}$, and $\Delta Z_H$. We can see that in both SLy5 and SLy5t cases, the mean value $\bar{A}_{H}$ of two channels are very close, while their standard deviations are quite different.
Specifically, the $\Delta A_H$ of the SI channel in SLy5t case is smaller than that of SLy5, while $\Delta A_H$ of the SII channel is larger than that of SLy5, which means that the heavy fragments near $135$ becomes concentrated and around 142 are more dispersed with the tensor force. 
Meanwhile, the $\bar{Z}_{H}$ and $\Delta Z_H$ present similar features.
Besides, for the SLy5 case, the primary fission fragments in the TDHF+FOA description of fission dynamics strongly depend on initial configurations. 
Namely, the initial configurations with larger quadrupole and octupole mass moments produce more heavy fission fragments.
A similar conclusion can also be drawn in SLy5t case.

To clarify the tensor effect on the internal structure of initial configurations, we apply the nucleon localization function (NLF) \cite{Reinhard2011_PRC83-034312,Zhang2016_PRC94-064323}, which is an excellent indicator of shell effects. 
The NLF is expressed as

\begin{equation}
	C_{q \sigma}(\bm{r})=\left[1+\left(\frac{\tau_{q \sigma} \rho_{q \sigma}-\frac{1}{4}\left|\nabla \rho_{q \sigma}\right|^{2}-\mathbf{j}_{q \sigma}^{2}}{\rho_{q \sigma} \tau_{q \sigma}^{\mathrm{TF}}}\right)^{2}\right]^{-1},
\end{equation}
where $q$ is the isospin ($n~\text{or}~p$) and $\sigma$ is the spin ($\uparrow~\text{or}~\downarrow$). $\tau^{\mathrm{TF}}=\frac{3}{5}\left(6 \pi^{2}\right)^{2 / 3} \rho_{q \sigma}^{5 / 3}$ is the Thomas–Fermi kinetic density and $\mathbf{j}$ is the current density. 
The NLF takes values between 0 and 1. When $C$ is close to one, it indicates that the probability of finding two particles (with the same isospin and spin) at spatial location $\bm{r}$ is very small. 
Thus, high values of $C$ show the shell structure. For a clear visualization of the localization inside the nucleus, the NLF can multiply a normalized particle density to suppress the localization far from the nucleus \cite{Zhang2016_PRC94-064323},
\begin{equation}
	C_{q \sigma}(\bm{r}) \rightarrow C_{q \sigma}(\bm{r}) \frac{\rho_{q \sigma}(\bm{r})}{\max \left[\rho_{q\sigma}(\bm{r})\right]}.
\end{equation}
In this work, the localization function is obtained by averaging over the spin $\sigma$, such as neutron localization function $C_{n}=(C_{n\uparrow}+C_{n\downarrow})/2$.

In Fig. \ref{Fig:Cp}, we display the NLF of protons $C_{p}$ of five configurations on the static fission pathway for SLy5 and SLy5t cases. These five configurations ($Q_{20}=80~\mathrm{b}, 88~\mathrm{b}, 97~\mathrm{b}, 106~\mathrm{b}, 114~\mathrm{b}$), as the initial states of dynamical evolution, are beyond the fission barrier.
For SLy5, the $C_{p}$ are presented in Figs. \ref{Fig:Cp} (a) - \ref{Fig:Cp} (e). 
It is seen that the shell structure of fragments is generally exhibited in all five configurations.
With the increase of $Q_{20}$, the neck becomes longer and thinner, and the rings of enhanced localization are gradually closed.
Interestingly, the $C_{p}$ in the neck region has a visible change when $Q_{20}$ increases to $106~\mathrm{b}$.
It shows that the octupole shapes within the fragments are strongly induced by the neck.
The resulting $C_{p}$ for SLy5t are displayed in Figs. \ref{Fig:Cp} (f) - \ref{Fig:Cp} (j).
With the tensor force, there is no significant difference of $C_{p}$ in the $Q_{20}=80~\mathrm{b}$ to $97~\mathrm{b}$. 
However, at larger elongations ($Q_{20}=106~\mathrm{b}$ and $114~\mathrm{b}$), the effect of the tensor force is to enhance the $C_{p}$ in the neck.
This enhancement indicates that nucleons become localized in the neck so that the pear-shaped fragments can be clearly seen here.
That is to say, after considering the tensor force, the shell structure of fragments in the initial configurations is enhanced.
Meanwhile, the neutron localization functions $C_{n}$ present similar features, as shown in Fig. \ref{Fig:Cn}.

\begin{figure*}[htbp]
	\includegraphics[width=0.75\textwidth]{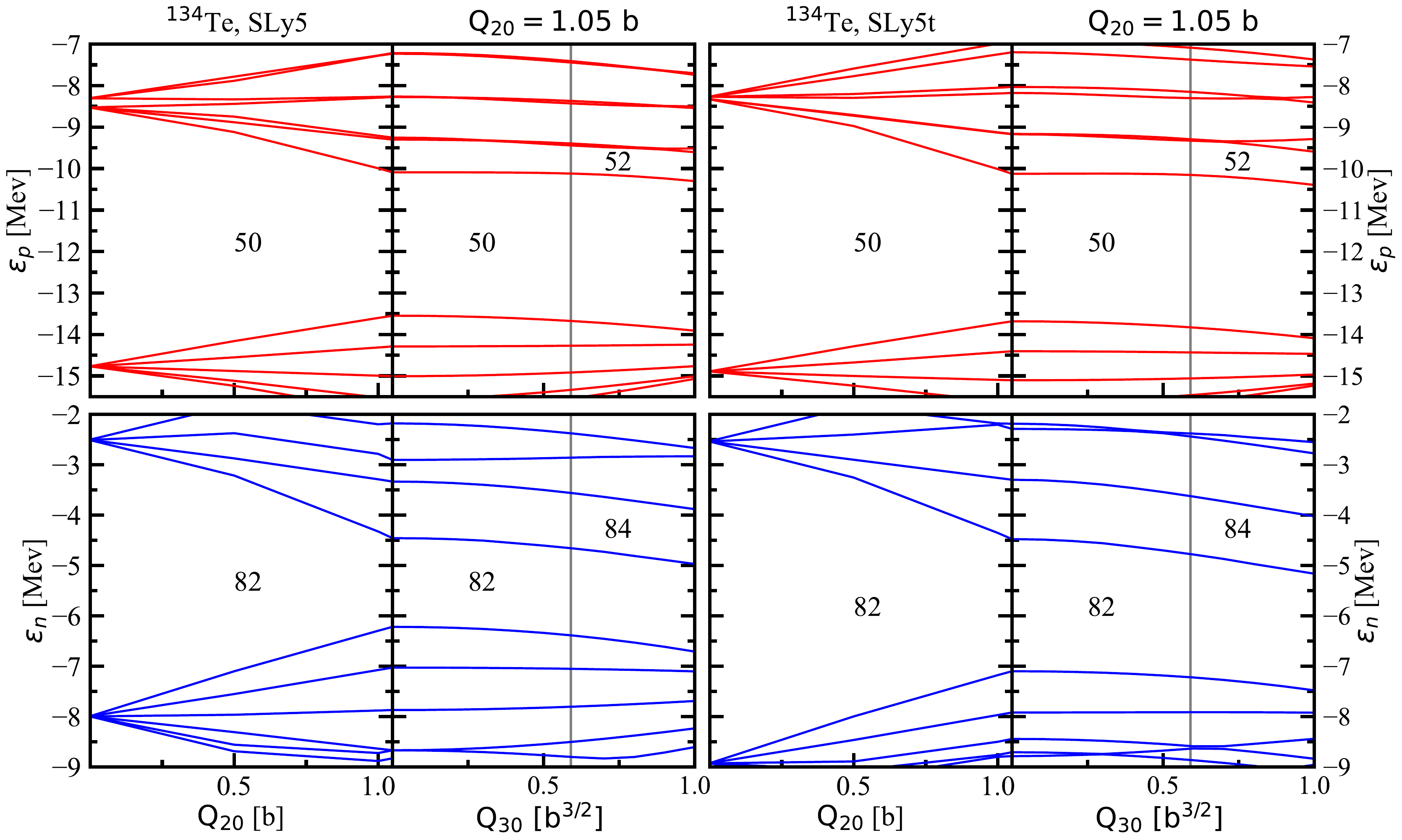}
	\caption{Evolution of single-particle levels with quadrupole mass moment $Q_{20}$ and octupole mass moment $Q_{30}$ for $^{134}\mathrm{Te}$ with SLy5 (the left panel) and SLy5t (the right panel) for protons (red line) and for neutrons (blue line). The gray vertical lines denote the $Q_{30}$ of the fission fragment. Shell gaps are indicated by the numbers 50, 52, 82, 84. }
	\label{Fig:Te134}
\end{figure*}

\begin{figure*}[htbp]
	\includegraphics[width=0.75\textwidth]{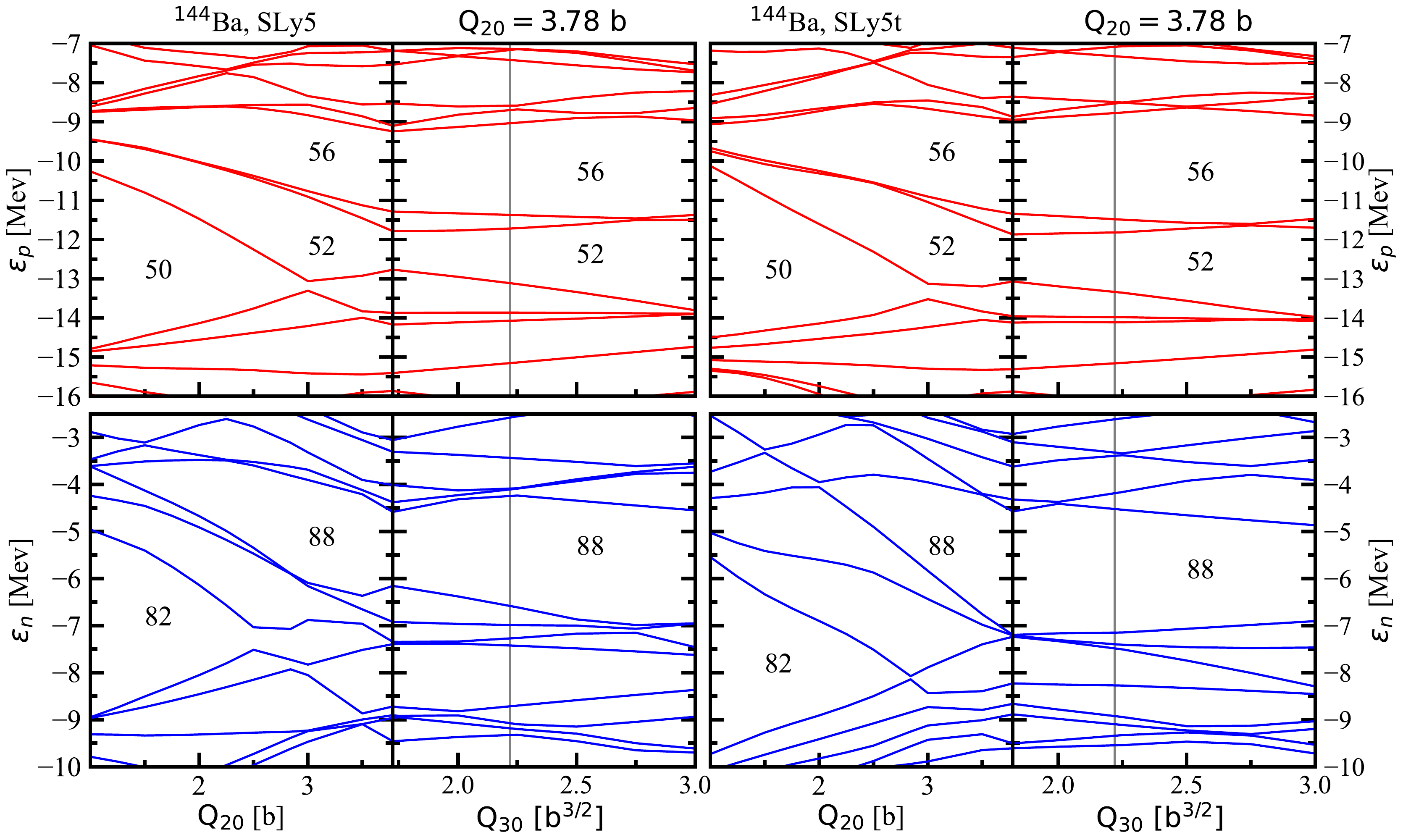}
	\caption{Similar to Fig. \ref{Fig:Te134} but for $^{144}\mathrm{Ba}$.}
	\label{Fig:Ba144}
\end{figure*}

Additionally, in Fig \ref{Fig:TimeEvolution}, one can also see that the bifurcation between the shape evolutions of two kinds of dynamical trajectories with SLy5t is more distinct than the case of SLy5. 
This phenomenon may be due to that the enhancement of shell gaps caused by the tensor force.
To understand this, we focus on the fission fragments and investigate the effect of tensor force on shell closures reflected by the structure of single-particle levels (SPLs), similar to the way shown in Ref. \cite{Scamps2018_Nature564-382}.
We select two representative fragments, the even-even nuclei $^{134}\mathrm{Te}$ and $^{144}\mathrm{Ba}$, which can be used to study the deformed shell gaps at $Z=52$ and $Z=56$, respectively.
The values of $Q_{20}$ and $Q_{30}$ of fragments at scission points are extracted from the wave function of dynamical simulation. 
Herein, we are interested in how the tensor force affects the evolution of deformed shell gaps.
Therefore, we constrain the shapes of these two fragments to be close to those at scission points to check the changes of SPLs.

Figure \ref{Fig:Te134} shows the evolution of proton and neutron single-particle energies $\epsilon_{p}$ (red lines) and $\epsilon_{n}$ (blue lines) with $Q_{20}$ and $Q_{30}$ for $^{134}\mathrm{Te}$. 
The left panel is obtained by SLy5 and the right panel for SLy5t. 
The left side of each panel shows the evolution of SPLs as an increase of $Q_{20}$ up to scission point.
The right side shows the evolution of SPLs with $Q_{30}$, where the value of $Q_{20}$ is fixed as the same as that at scission point.
For $^{134}\mathrm{Te}$, The quadrupole and octupole mass moments at scission point are $Q_{20}\approx1.05~\mathrm{b}$ (shown at the top of panel) and $Q_{30}\approx0.59~\mathrm{b}^{3/2}$ (labeled by the gray vertical lines).
In both SLy5 and SLy5t cases, the large shell gaps at $Z=50$ and $N=82$ play the dominant role for the formation of fragments.
In comparison with $Z=50$, the deformed shell gap $Z=52$ is relatively small.
This is due to that the shell gaps at Z=50 and N=82 are not influenced too much since the deformation at the scission point is small.
With the inclusion of tensor force, the gaps at $Z=50$ and $Z=52$ are almost unchanged, while for neutrons, there is a strong enhancement of the $N=82$ gap. 
For $^{144}\mathrm{Ba}$, the evolution of single-particle energies $\epsilon_{p}$ (red lines) and $\epsilon_{n}$ (blue lines) is shown in Fig. \ref{Fig:Ba144}.
At scission point, the associated deformations of $^{144}\mathrm{Ba}$ are $Q_{20}\approx3.78~\mathrm{b}$ (shown at the top of panel) and $Q_{30}\approx2.22~\mathrm{b}^{3/2}$ (labeled by gray vertical lines).
The $^{144}\mathrm{Ba}$ exhibits the octupole deformation in its ground state, which has been confirmed experimentally \cite{Bucher2016_PRL116-112503}.
For SPLs of protons, as the $Q_{20}$ increases, the spherical shell $Z=50$ is quenched and the deformed proton shell gaps at $Z=52$ and $Z=56$ gradually play a role. 
When the $Q_{20}$ is fixed to be $3.78~\mathrm{b}$, the $Z=52$ and $Z=56$ gaps are retained along with the function of $Q_{30}$.
By including the tensor force, a slight enhancement of the $Z=56$ shell gap at the larger $Q_{30}$ is found.
For SPLs of neutrons, it is clear that the spherical shell gap at $N=82$ disappears due to deformation effects and the shell closure at $N=88$ plays a dominant role. 
The inclusion of the tensor force also slightly enhances the $N=88$ shell gap, especially at $Q_{30} \leq 2.22~\mathrm{b}^{3/2}$.
These enhancements of the deformed shell gaps give a microscopic interpretation of the distinct bifurcation between the SI and SII channels.
Overall, we demonstrate that the tensor force reinforces the shell effect of fission fragments.

\begin{figure}[htbp]
	\includegraphics[width=0.45\textwidth]{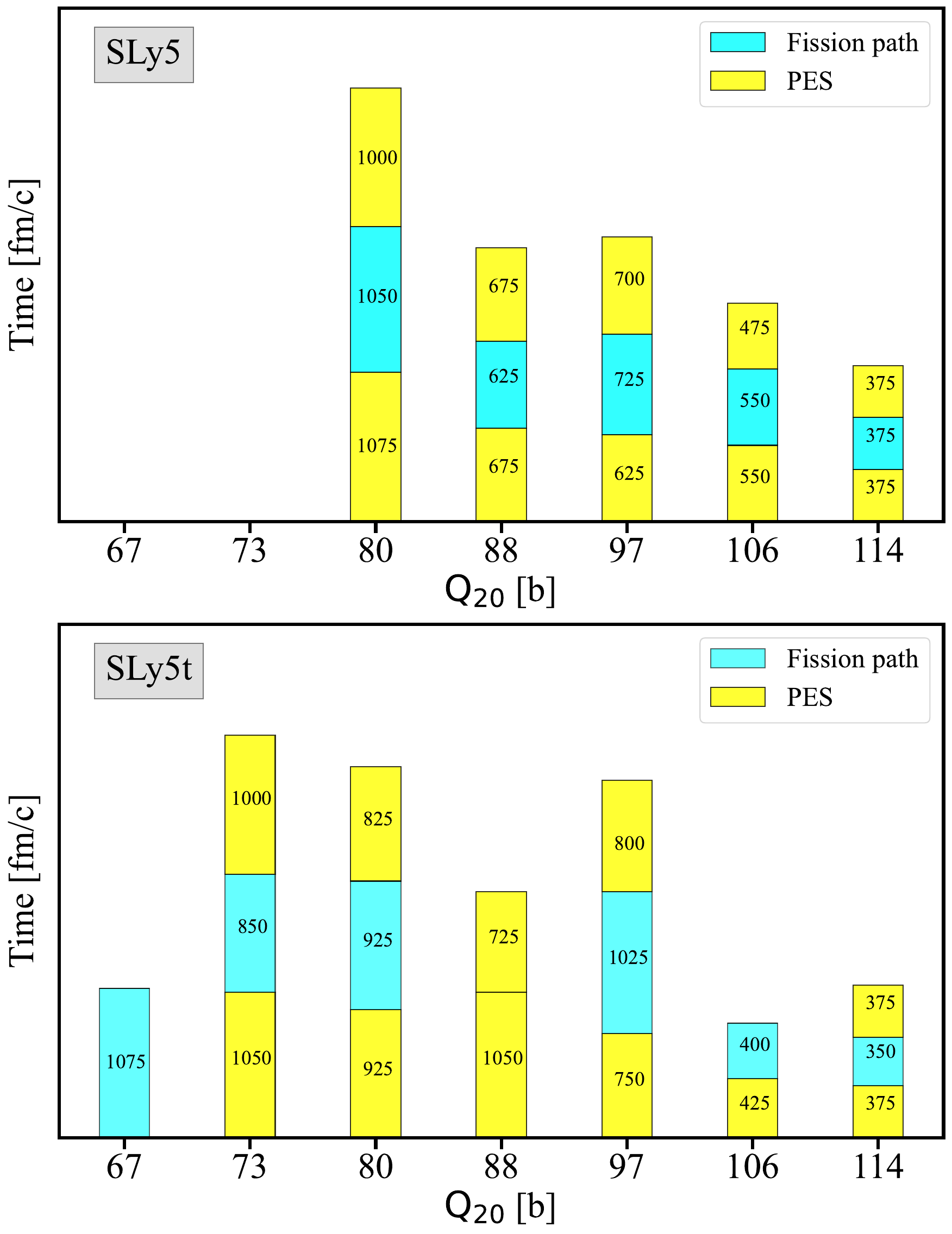}
	\caption{The time scale of SLy5 (the upper panel) and SLy5t (the bottom panel) for various static configurations. 
	The cyan bars present the initial points taken from the static fission path and the yellow bars show the rest of the points taken from the PES.
		The specific numerical values are also shown. }
	\label{Fig:FissionTime}
\end{figure}

\begin{figure}[htbp]
	\includegraphics[width=0.45\textwidth]{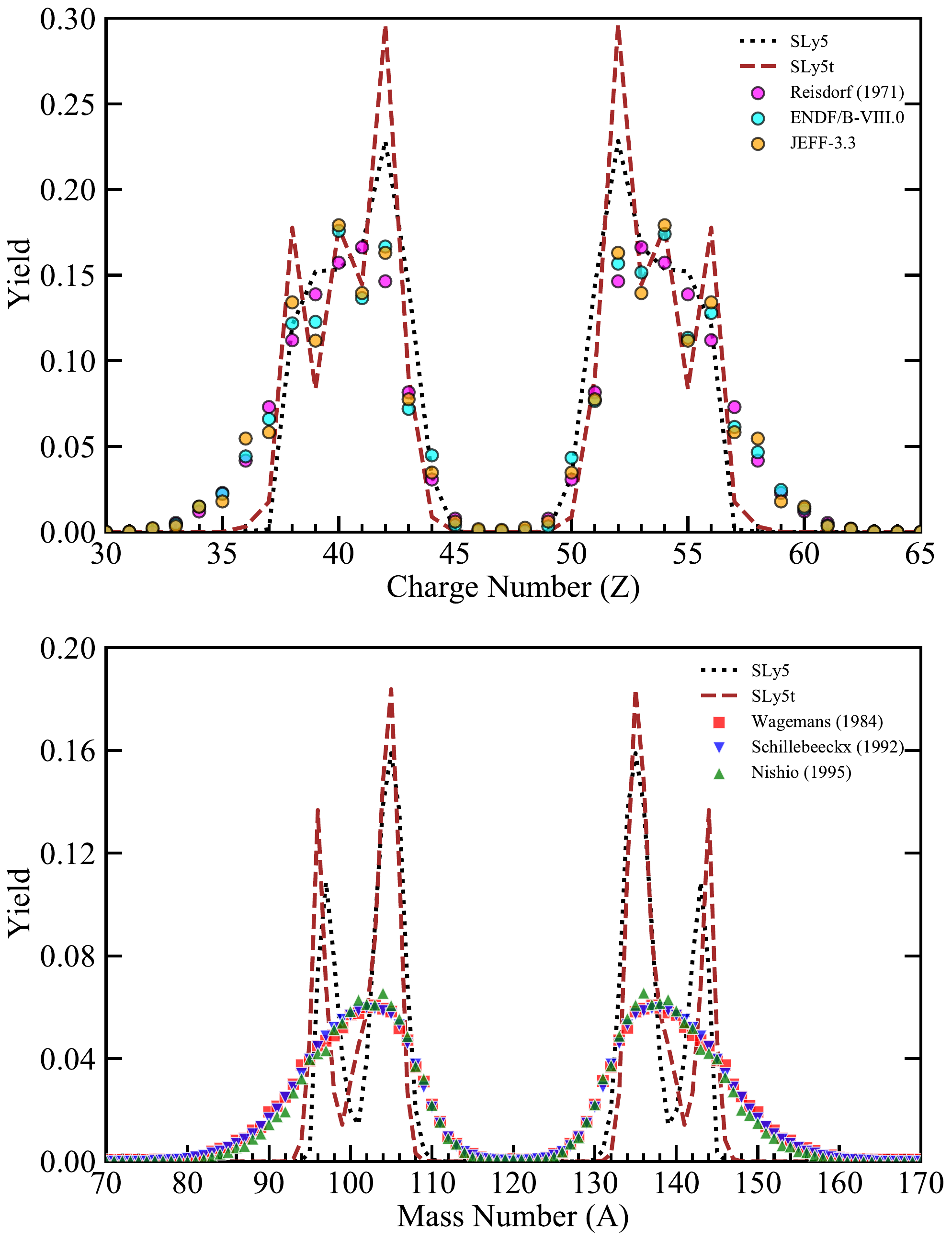}
	\caption{The distribution of charge (the upper panel) and mass (the bottom panel) for the deformation-induced fission of $^{240}\mathrm{Pu}$. 
	The lines represent the theoretical calculations using the double PNP.
	The experimental data taken from Wagemans 1984 \cite{Wagemans1984_PRC30-218}, Schillebeeckx 1992 \cite{Schillebeeckx1992_NPA545-623}, Nishio 1995 \cite{Nishio1995_JNST32-404}, Reisdorf 1971 \cite{Reisdorf1971_NPA177-337}, ENDF/B-VIII.0 library \cite{Brown2018_NDS148-1} and JEFF-3.3 library \cite{Plompen2020_EPJA56-181} are shown for comparison.}
	\label{Fig:rawYield}
\end{figure}

\begin{figure}[htbp]
	\includegraphics[width=0.45\textwidth]{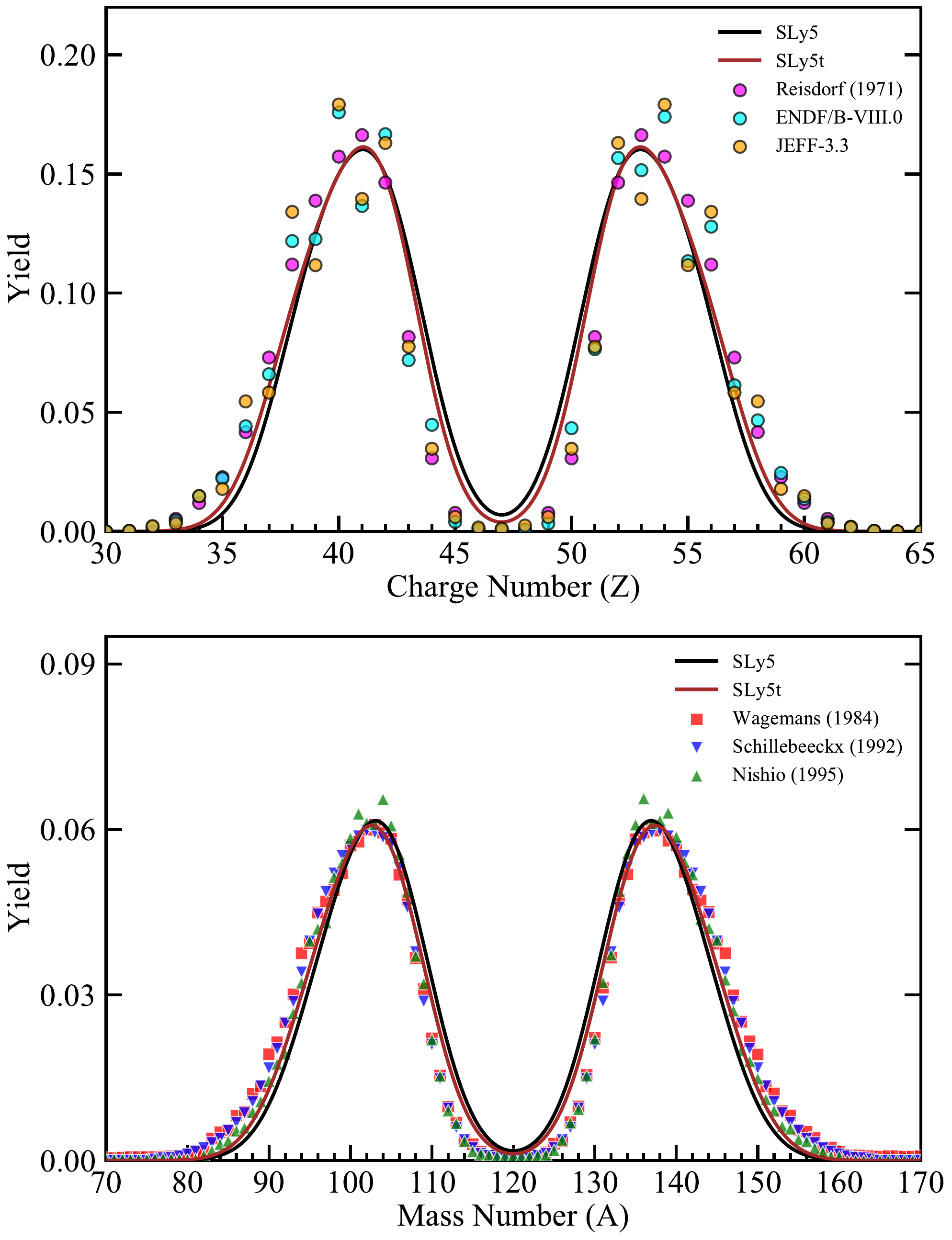}
	\caption{The distribution of charge (the upper panel) and mass (the bottom panel) for the deformation-induced fission of $^{240}\mathrm{Pu}$. 
	The solid lines show the results using the combination of double PNP and Gaussian kernel estimation with $\sigma=5.1$ and 1.6 for mass and charge distributions.
	The same experimental data as Fig. \ref{Fig:rawYield} are shown.}
	\label{Fig:distributions}
\end{figure}

Connected to the evolution of shape, the time scale is one of the intriguing questions about fission dynamics. 
In induced fission, we focus on the time of the collective motion from the initial configuration to the scission point, which strongly depends on the characteristics of the PES and plays an important role in the TKE of fragments. 
We use the definition of scission point in Ref. \cite{Goddard2015_PRC92-054610} and in the present work, the minimum density between the centers of two fragments is set to be 0.03 $\mathrm{fm}^{-3}$.
In Fig. \ref{Fig:FissionTime}, we present the time to scission points for all fissioning states.
For the initial points on the static fission path of SLy5 (marked by cyan bars), starting from $80~\mathrm{b}$ to $114~\mathrm{b}$, the time scale along with the quadrupole mass moment $Q_{20}$ has a decreasing tendency, ranging from $1050~\mathrm{fm/c}$ to $375~\mathrm{fm/c}$. 
It means the bigger deformation, the less time scale. 
In SLy5t case, the situation becomes complex as there is a non-fissioning state on the static fission path. 
Focusing on the cyan bars in Fig. \ref{Fig:FissionTime}, as the $Q_{20}$ increases, the time scale decreases gradually at first, then increases when the $Q_{20}$ beyond the non-fissioning state ($Q_{20}=88~\mathrm{b}$), and finally decreases rapidly.
The time scale required at $Q_{20}=97~\mathrm{b}$ increases up to $1025~\mathrm{fm/c}$ which is comparable to the initial point with $Q_{20}=67~\mathrm{b}$. 
As pointed out in Ref. \cite{Scamps2015_PRC92-011602}, the times to reach scission may be dependent on the slope of PES and the one-body viscosity.
Hence, the long time scale of $Q_{20}=97~\mathrm{b}$ might be caused by the special behavior of the octupole mass moment, as mentioned before.
Moreover, according to the time scale of those states that are not located at static fission path, there is no obvious regularity in the time scales of different octupole moments.

After investigating the tensor effect on fission dynamics, we focus on its effect on the mass and charge distributions of fission fragments of $^{240}$Pu. 
The double PNP [cf. Eq. (\ref{2})] is applied to the fission outcomes of dynamical calculation. 
In this work, we assume equal weights for all fission events.
It is supposed to account for three factors: (i) the initial points are chosen from the fission valley in a nearly equidistant pattern in both $Q_{20}$ and $Q_{30}$ directions, ensuring that the dominant fission events could be involved, (ii) all the initial configurations are equally important as each fission event in TDDFT simulations is independent \cite{Goddard2015_PRC92-054610,Scamps2018_Nature564-382}, and (iii) a recently work combining the TDDFT and TDGCM has shown that the trajectories with initial deformations close to the static fission pathway play a dominated role while the remaining ones are not significant \cite{Ren2022_PRC105-044313}.
Hence, assuming that all trajectories have the same weight is desirable and reasonable and it is more convenient to study the effect of the tensor force.
Finally, the distributions of fission fragments can be calculated by directly summing up the projected probabilities for all fission trajectories. 

The mass and charge distributions from TDHF$+$FOA with double PNP calculations for SLy5 and SLy5t are presented and
compared with experimental data in Fig. \ref{Fig:rawYield}.
As for the charge distributions, 
the width of SLy5t is consistent with that of SLy5 and also with the experiments. 
There is the same peak at $Z=52$ in SLy5 and SLy5t, which is originated from the SI channel. However, for the SII channel, the structures of the charge distribution of SLy5 and SLy5t are different. In SLy5 case (black dotted lines), the charge distribution does not have a peak at $Z=56$, but there is a small ``plateau'' consisting of the similar probability of $Z=53-56$. This feature seems to provide a clue to the occurrence of odd-even effects. 
After including the tensor force (brown dashed lines), one can see that there are two additional peaks in the distribution for heavy fragments at $Z=54$ and $Z=56$. Specifically, the yields of even-$Z$ fragments are enhanced compared with odd-$Z$ fragments.
In other words, we observed a clear odd-even effect in the charge distribution of SLy5t.
A direct insight into these results can be obtained from $\bar{Z}_{H}$ and $\Delta Z_H$, see Table \ref{table:meanvalue}. 
The SI channel of SLy5 and SLy5t have $\bar{Z}_{H}\approx 52$, and their $\Delta Z_H$ are both small, not greater than 0.5, which means that the contribution of distribution is mainly around at $Z=52$. 
However, the SII channel of SLy5 and SLy5t have the same $\bar{Z}_{H}$, but their $\Delta Z_H$ are very different. The $\Delta Z_H$ of SLy5 is only 0.36, and about 1 of SLy5t, which means that the contribution from $Z=54$ and $Z=56$ becomes larger after including tensor force.
	Furthermore, a more microscopic explanation is given by the enhancement of the tensor force on the deformed proton shells $Z=52$ and $Z=56$ in the heavy fission fragments. 
	Strong proton shells observed in fragments favor the occurrence of these proton numbers.
	That is to say, the probability of $Z=52$ or $56$ in a single fission event increases with the inclusion of the tensor force.
	Hence, after considering all fission events, the odd-even effect could be shown in the final charge distribution.
By the way, we do not observe the odd-even effect in the distribution of each single fission event in this work.
It is not surprising because the presence of such an effect in single events is strongly related to the scission configuration \cite{Verriere2019_PRC100-024612,Verriere2021_PRC103-034617,Verriere2021_PRC103-054602}; that is, only specific cases will show this structure.
However, in the TDDFT framework, the configurations of scission point are automatically generated 
during non-adiabatic dynamics rather than artificially selected.

\begin{figure}[htbp]
	\includegraphics[width=0.48\textwidth]{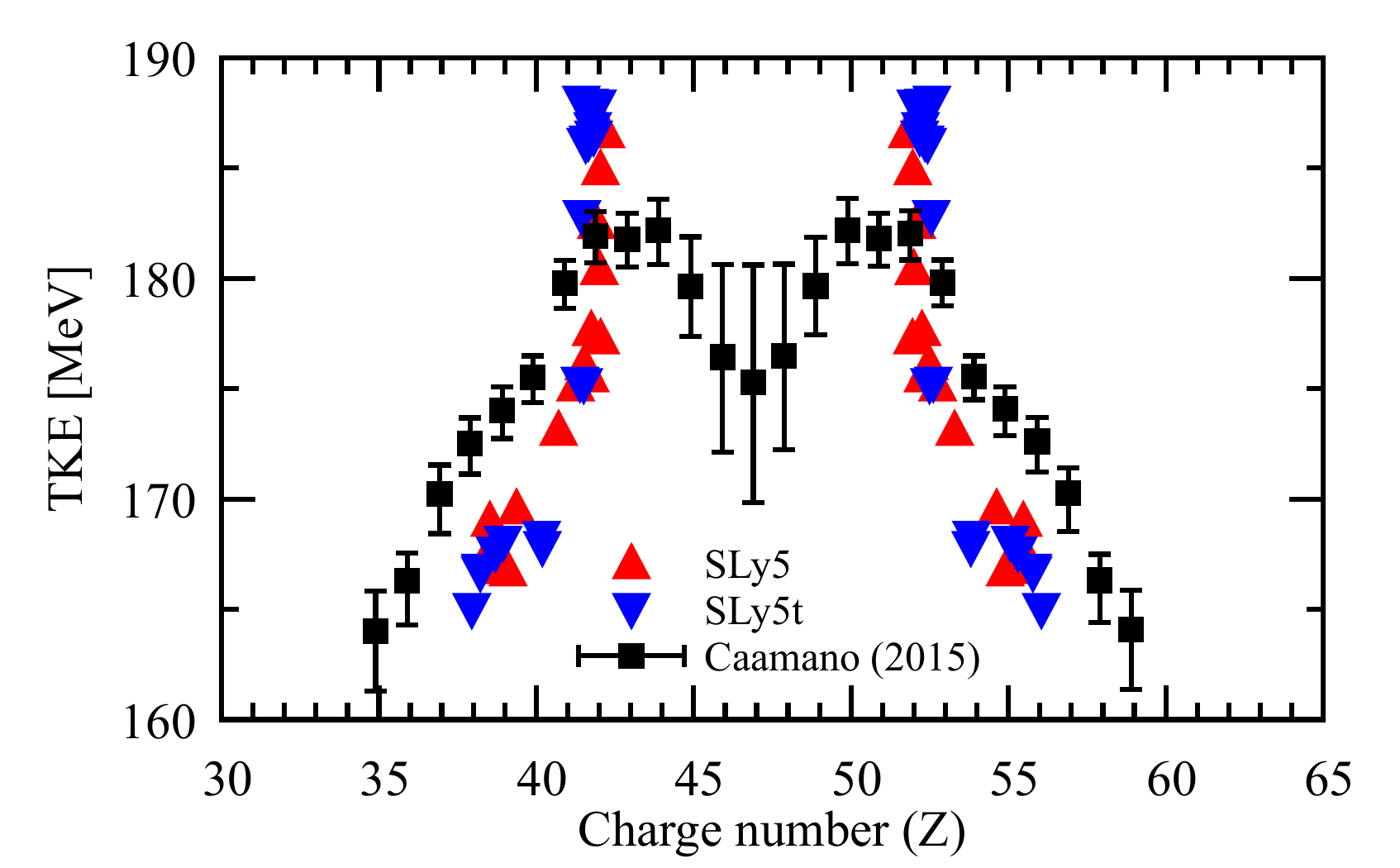}
	\caption{The total kinetic energies of the fission fragments of $^{240}\mathrm{Pu}$ as a function of the fragment's charge, calculated with effective interactions SLy5 and SLy5t. The experimental data taken from Caama\~no 2015 \cite{Caamano2015_PRC92-034606} are shown for comparison.}
	\label{Fig:Z-TKE-2D}
\end{figure}

For the mass distributions, two peaks appear in both SLy5 and SLy5t cases, corresponding to the SI and SII fission channels respectively. 
After including the tensor force, the peak at $A_H = 135$ does not change. This peak corresponds to the maximum yield for heavy fragments, which is consistent with the peak value given by the experimental data.  
The other peak slightly shifts to the asymmetric part of distribution, from $A_H = 143$ to $A_H = 144$.
In addition, for both cases, one can see an obvious drop between the two peaks, which means a rapid change from SI channel to SII channel. 
Obviously, the mass distributions calculated by SLy5 and SLy5t do not agree with the experimental mass distribution.
In particular, there is a notable difference discrepancy in the distribution width.
The insufficient incorporation of quantum fluctuations in the adiabatic processes is the main reason 
that the mass distributions does not reproduce well, which has been discussed in detail in our previous work \cite{Huang2024_EPJA60-100}. 
We expect that combining TDBCS, double PNP and a reasonable probability for fission trajectories will be an advanced way to consider the quantum fluctuations in fission dynamics and thus provide a proper prediction of the distributions of fission fragments. 
But this will be a long-term project.
Hence, in this work, we follow the same route as our previous study \cite{Huang2024_EPJA60-100} and focus on discussing the impact of tensor forces on the distributions.
The Gaussian kernel estimation (GKE) is used to compensate for the lack of quantum fluctuations. 
The set of Gaussian width is $\sigma=5.1$ for mass yield and $\sigma=1.6$ for charge yield, which has been determined by investigating how GKE influences the distributions from TDHF$+$FOA$+$PNP calculations.
Herein, we take the same width set for both SLy5 and SLy5t cases. 

Figure \ref{Fig:distributions} shows the calculated mass and charge distributions using GKE based on the distribution from TDHF$+$FOA$+$PNP for SLy5 and SLy5t, along with the comparison with the experiments.
	Our results show that both the positions of asymmetric pecks and the widths of mass distributions are slightly improved when the tensor force is involved (shown by the brown line). 
	Besides, we notice a reduction in symmetric yields and an enhancement of more asymmetric yields for SLy5t, which improves the results and makes them more consistent with experimental data.
	This is due to that the deformed shells of heavy fragments are reinforced by the tensor force, resulting in a shift of the mass distribution towards larger values of $A_{H}$.
	Similarly, the inclusion of tensor force also improves the descriptions of charge distribution, as seen in Fig. \ref{Fig:distributions}.
	However, the odd-even effect in charge distribution for SLy5t calculations has disappeared due to the smoothing influence from the GKE. 
	The phenomenological GKE is typically used to take the number of nucleons in the pre-fragments and the experimental resolution into account within the adiabatic TDGCM framework.
	As far as we know, even the TDGCM$+$PNP cannot reproduce such effects \cite{Verriere2021_PRC103-054602}.
	Finally, we would like to mention that TDHF+FOA+PNP with GKE is an alternative way for predicting fission distribution, rather than a fully microscopic approach.
	Further study of the influence of quantum fluctuations using microscopic methods is interesting and significant.

Apart from the mass and charge distributions, the TKE of fission fragments is another important quantity, which directly relates to the occurrence of scission point, the evolution of dynamical process, and the fission models.
In the TDHF theory, TKE is an asymptotic value obtained by summing the nuclear collective kinetic energy and the Coulomb energy at a relatively large distance between two fission fragments ($\approx30$~fm in our calculations) \cite{Goddard2015_PRC92-054610} and reads
\begin{equation}
	E_{\mathrm{TKE}}=E_{\mathrm{coll.kin.}}+E_{\mathrm{Coul}},
\end{equation}
where the collective kinetic energy is defined as 
\begin{equation}
E_{\text {coll.kin.}}=\frac{\hbar^{2}}{2 m} \int d \bm{r} \frac{\mathbf{j}(\bm{r})^{2}}{\rho(\bm{r})},
\end{equation}
and $\mathbf{j}(\bm{r})$ is the total current density. 
The TKEs of the nascent fission fragments for $^{240}\mathrm{Pu}$ with SLy5 and SLy5t as functions of the fragment charge number are shown in Fig. \ref{Fig:Z-TKE-2D}. 
The experimental data taken from Ref. \cite{Caamano2015_PRC92-034606} is also shown for comparison.
In general, theoretical results for both cases are distributed around the experimental data. 
The peaks of the TKEs, which are associated with a compact configuration at the scission point, are higher than experiments, and the tails of the TKEs, which correspond to a more elongated, asymmetric shape, are lower than experiments. 
These features are likely to be due to two factors: the initial configurations \cite{Goddard2015_PRC92-054610}, and the dynamical effects such as the one-body dissipative effect \cite{Bulgac2019_PRC100-034615,Ren2022_PRC105-044313}. 
That is to say, it is related to the energy transformation between adiabatic and non-adiabatic processes.
After the tensor force is taken into account, the distribution of TKEs shows a slight increase at $Z=52$ and a decrease at $Z=56$. 
It is intriguing to note that the calculated TKEs exhibit a clear focusing phenomenon despite the fact that initial configurations are selected from a 2D PES. 
These results are related to the distinct bifurcation observed in the shape evolutions of SI and SII dynamical trajectories after considering the tensor force.

\section{Summary and perspective}
\label{summary}
In this work, the effect of the tensor force on fission dynamics has been studied.
Combining the constrained HF+BCS, TDHF+FOA with double particle number projection, and the Gaussian kernel estimation, we investigate the static fission pathway, the dynamical evolution from saddle to scission points, the mass and charge distributions and total kinetic energies of fission fragments in the induced fission of $^{240}\mathrm{Pu}$.
The 2D PES in the ($Q_{20},~Q_{30}$) plane of $^{240}\mathrm{Pu}$ is calculated by using the CHF+BCS with the SLy5 and SLy5t effective interactions. 
By comparing the static fission pathway and the PES obtained from SLy5 and SLy5t calculations, we find that the tensor force influences both the inner and outer barrier height, and makes the double-humped structure of static fission path more consistent with empirical one.
Meanwhile, the tensor force renders the region of the fission valley wider and suppresses the triaxial deformation around the outer barrier.
Next, the fission dynamics are simulated by using TDHF with the frozen occupation approximation. 
It is interesting that the tensor force enhances the difference in shape evolution between the SI and SII channels. 
Its microscopic interpretation is that the tensor force strengthens the deformed shell effects in fission fragments.
Then, the double PNP is used to calculate the mass and charge distributions of fission fragments for each single fission trajectory.
The TDHF+FOA+PNP method can well repoduced the charge distributions in both SLy5 and SLy5t cases. 
An interesting result is that the odd-even effect appears in the charge distribution when the inclusion of the tensor force.
After introducing the GKE, the mass and charge distributions can be simultaneously in good agreement with the experiments.
Our calculations show that the inclusion of tensor force leads to an increase in more asymmetric fragments of mass and charge distributions, and then to a better agreement with experimental data. 
Additionally, the calculated TKE of fission fragments is also reasonably consistent with the experiments, and the inclusion of tensor force revealed a focusing phenomenon that supports the behavior of shape evolution.
In summary, we conclude that the tensor force strongly affects the PES, as well as the dynamical process. 

In our present approach, 
the pairing treatment of fission dynamics is the BCS at the FOA level, in which the occupation numbers are frozen during the time evolution.
This approach leads to the threshold anomaly, and then this problem becomes more troublesome when initial states are selected from a multi-dimensional PES.
One should use a more general method for treating the dynamical pairing to reduce the effect of the threshold anomaly, such as the time-dependent BCS (TDBCS) or the Hartree-Fock-Bogoliubov (TDHFB).
Furthermore, the adiabatic fluctuations associated with each fission trajectory are crucial for calculating the fission fragment distributions.
A more appropriate method for accounting for the weight of initial states in TDDFT might be the WKB approximation or Markov chain Monte Carlo sampling.
Considering more adiabatic fluctuations can improve the prediction power of theoretical models for the distributions of fission fragments. 
Including these improvements in the TDDFT framework is our next goal, and thus, it will be crucial to study the odd-even effect of the experimental charge yield, and even the asymmetric to symmetric transition in the mass distribution of light thorium isotopes observed experimentally \cite{Chatillon2019_PRC99-054628,Chatillon2020_PRL124-202502}.
Moreover, it will also be interesting to see the effect of tensor force on other aspects of fission dynamics, such as the generation of angular momentum \cite{Wilson2021_Nature590-566}. 
Finally, further investigation of its effect on the other actinides, or the sub-lead region, will broaden our understanding of the microscopic mechanisms entailed in fission.

\begin{acknowledgments}
We thank Liang Li and Zhen-Ji Wu for helpful suggestions. 
This work has been supported by the National Natural Science Foundation of China (Grants No. 12375127, No. 11975237, and No. 12175151),
the Strategic Priority Research Program of Chinese Academy of Sciences under Grant No. XDB34010000, and the Fundamental Research Funds for the Central Universities under Grant No. E2E46302.
The results described in this work are obtained on the C3S2 computing center of Huzhou University.
Xiang-Xiang Sun is supported in part by 
NSFC under Grants No. 12205308, and the Deutsche Forschungsgemeinschaft
(DFG) and NSFC through the  funds provided to the Sino-German Collaborative Research Center TRR110  ``Symmetries and the Emergence of  Structure in QCD''
(NSFC Grant No. 12070131001, DFG Project-ID 196253076).
\end{acknowledgments}

\bibliographystyle{apsrev4-2}
\bibliography{ref}
\end{document}